\documentclass[]{aastex631}

\usepackage{natbib}
\bibpunct{(}{)}{;}{a}{}{,} 

\usepackage{graphicx}
\usepackage{bm}

\usepackage{amsmath}
\usepackage{tabularx}
\usepackage{multirow}

\usepackage{savesym}
\savesymbol{tablenum}
\usepackage{siunitx}
\restoresymbol{SIX}{tablenum}

\usepackage{placeins}
\usepackage{hyperref}
\usepackage{ulem}

\DeclareSIUnit[]\rsun
{\text{R\ensuremath{_{\textup{sun}}}}}

\DeclareMathOperator{\sinc}{sinc}
\usepackage{booktabs} 

\begin{document} 

    \title{Revised Point-Spread Functions for the Atmospheric Imaging Assembly onboard the Solar Dynamics Observatory}

    \author[0000-0001-7662-1960]{Stefan J. Hofmeister}
    \affiliation{Columbia Astrophysics Laboratory, Columbia University, 538 West 120th Street, New York, NY 10027}
  
    \author[0000-0002-1111-6610]{Daniel W. Savin}
    \affiliation{Columbia Astrophysics Laboratory, Columbia University, 538 West 120th Street, New York, NY 10027}
   
    \author[0000-0001-7748-4179]{Michael Hahn}
    \affiliation{Columbia Astrophysics Laboratory, Columbia University, 538 West 120th Street, New York, NY 10027}

   \date{\today}

  \begin{abstract}
We present revised point-spread functions (PSFs) for the Atmospheric Imaging Assembly (AIA) onboard the Solar Dynamics Observatory (SDO). These PSFs provide a robust estimate of the light diffracted by the meshes holding the entrance and focal plane filters and the light that is diffusely scattered over the detector by the micro-roughness of the mirrors. 
We first calibrate the diffracted light using flare images. Our modeling of the diffracted light provides reliable determinations of the mesh parameters and finds that about $24$~to \SI{33}{\percent} of the collected light is diffracted, depending on the AIA channel. Then, we fit the diffuse scattered light using partially lunar occulted images. We find that the diffuse scattered light can be modeled as a superposition of two power law functions that scatter light over the entire detector. The amount of diffuse scattered light ranges from $10$~to \SI{35}{\percent}, depending on the AIA channel. In total, AIA diffracts and diffusely scatters about $37$~to \SI{55}{\percent} of the collected light over the detector.
When correcting for this, bright image regions increase in intensity by about \SI{30}{\percent}, dark image regions decrease by up to \SI{90}{\percent}, and the associated differential emission measure analysis of solar features is affected accordingly. Finally, we compare the image reconstructions using our new PSFs to those from the AIA team and \citet{poduval2013}. We find that our PSFs outperform the others, better correcting for the flare diffraction pattern and far more accurately predicting long-distance scattered light in lunar occultations.

   \end{abstract}


\section{Introduction}

The point-spread function (PSF) of an imaging system describes the optical aberration of the system. The PSF corresponds to the observed intensity distribution of a point source imaged onto the detector. Short-distance scattered light primarily results in blurring, while medium- to long-distance scattered light primarily reduces the dynamic contrast in the collected images. By using the PSF to correct the collected images, one can reconstruct the true images, i.e., the images without instrumental aberrations \citep{zhang2016}.  

The PSF of an imaging system is given by the convolution of the PSFs of all relevant optical components, such as lenses, mirrors, meshes, gratings, and field stops. However, the PSFs of the individual components are generally not known with sufficient accuracy. Instead, a more pragmatic approach is taken, where one derives the combined PSF directly from certain types of special images, such as bright point sources or partially occulted images \citep{grigis2012, poduval2013, hofmeister2022}. 

The Atmospheric Imaging Assembly (AIA) onboard the Solar Dynamics Observatory (SDO) observes the Sun in seven extreme ultraviolet (EUV) channels. The PSF of each channel has three main contributions: (1) a PSF core, which primarily results in blurring; (2) a complex diffraction pattern from the meshes that support the entrance and focal plane filters, which result in directionally scattered light over medium-to-long distances; and (3) diffuse scattered light components, which scatter light over the entire detector, reduce the dynamic contrast of the images, and probably arises from the micro-roughness of the mirrors. Medium-distance scattered light is commonly referred to as the PSF wing and long-distance scattered light as the PSF tail. We define the PSF core as the central pixel of the PSF and its adjacent pixels, the PSF wing as light scattered \SI{\le 10}{px} from the central pixel, and the PSF tail as light scattered \SI{>10}{px} away.

\begin{figure}
    \centering
    \includegraphics[width=\textwidth]{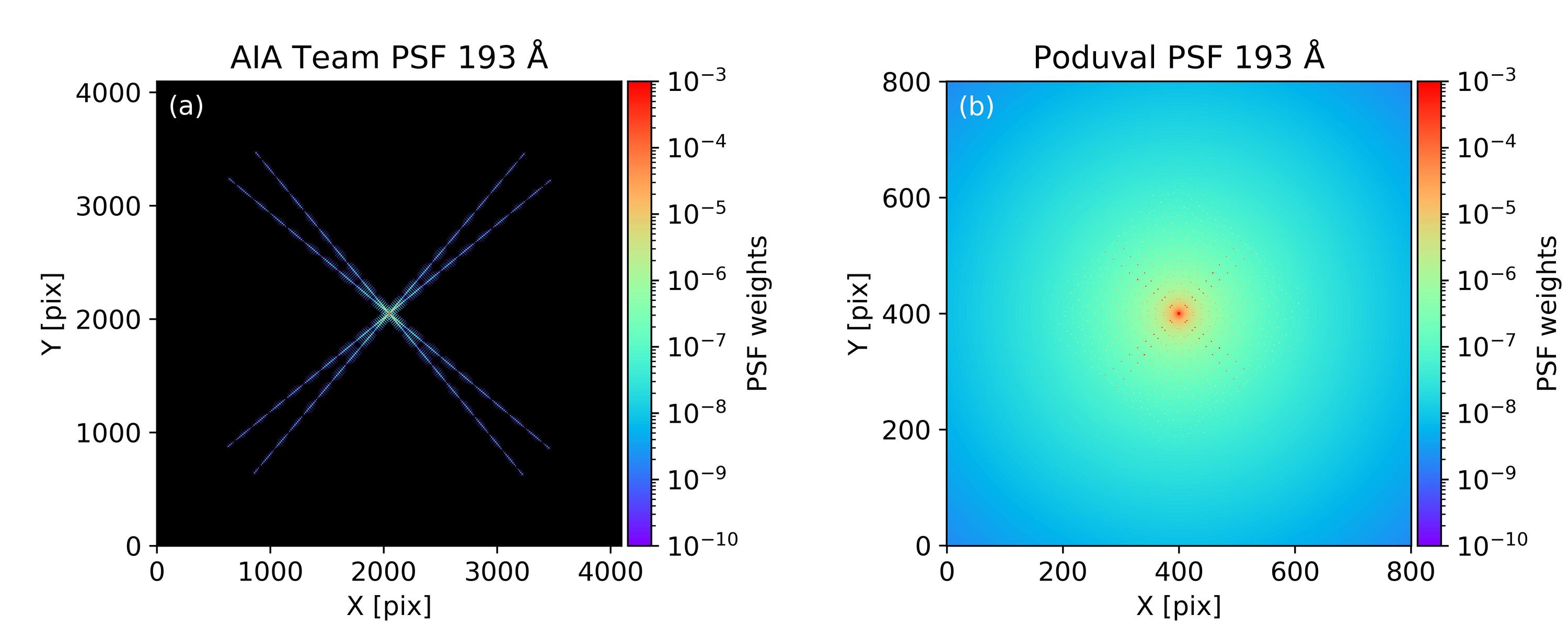}
    \caption{AIA PSF of the \SI{193}{\angstrom} channel as derived (a) by the AIA instrument team \citep{grigis2012} and (b) by \citet{poduval2013}. The PSF of the instrument team has  a size of $4096 \times 4096$ pixels and consists of a PSF core, two diffraction patterns arising from two meshes that support the entrance filter, and one diffraction pattern from a mesh that supports the focal plane filter. The image only shows the diffraction patterns from the meshes supporting the entrance filter; the diffraction pattern of the mesh supporting the focal plane filter is very weak and cannot be seen here. The PSF derived by \citet{poduval2013} has a size of $800 \times 800$ pixels, consists of the PSF core, the diffraction patterns, and a diffuse scattered light component. }
    \label{fig:overview}
\end{figure}

\begin{figure}
    \centering
    \includegraphics[width=\textwidth]{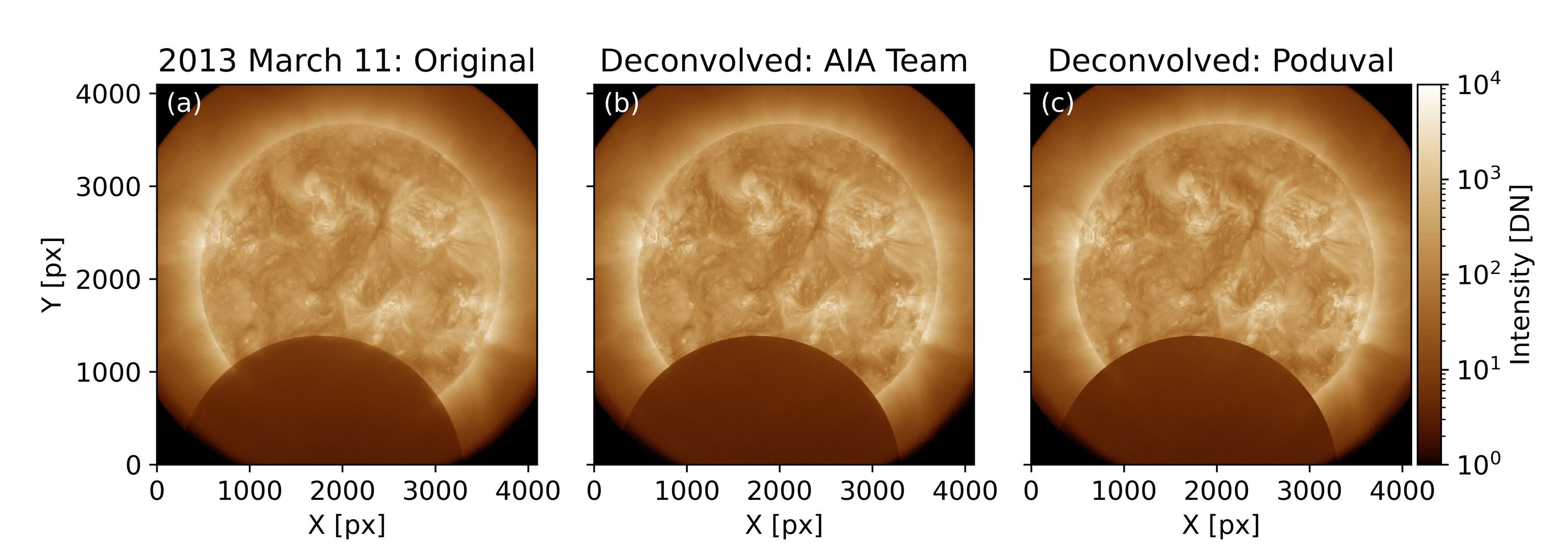}
    \caption{Lunar transit images taken by AIA in the \SI{193}{\angstrom} channel on 2013~March~11. (a) Original image. (b) Deconvolved image using the PSF of the AIA instrument team \citep{grigis2012}. (c) Deconvolved image using the PSF of \citet{poduval2013}. Both deconvolved images still show a clear residual intensity profile in the lunar-occulted pixels.}
    \label{fig:overview2}
\end{figure}

The first estimations of the PSF were provided by the AIA instrument team \citep{grigis2012} and considered only the PSF core and the diffraction pattern. Subsequently, \citet{poduval2013} tried to fit the diffuse scattered light component by evaluating the residual intensities in the occulted pixels during a Venus transit. These PSFs are shown in Figures~\ref{fig:overview}(a) and~(b), respectively. However, lunar transits show that the long-distance diffuse scattered light is still underestimated. The PSF reconstructed intensities in the lunar-occulted pixels should be zero; but instead, they still show a residual intensity profile (Figure~\ref{fig:overview2}). 

In this study, we have re-calibrated the PSFs of AIA by employing flare and lunar-transit data collected over the years 2010 to~2023. We have re-calibrated the diffraction pattern, fit the diffuse scattered light, and analyzed the PSF core. We have then evaluated how the revised PSFs have improved the quality of the reconstructed images. Our study also aims to serve as a guideline for future PSF calibrations for EUV imagers in solar physics. To this end, we also discuss the most important issues we have encountered during the PSF calibration and how we have addressed these issues.

\section{The AIA instrument} 
\label{sec:aiaoverview}


\begin{figure}
    \centering
    \includegraphics[width=\textwidth]{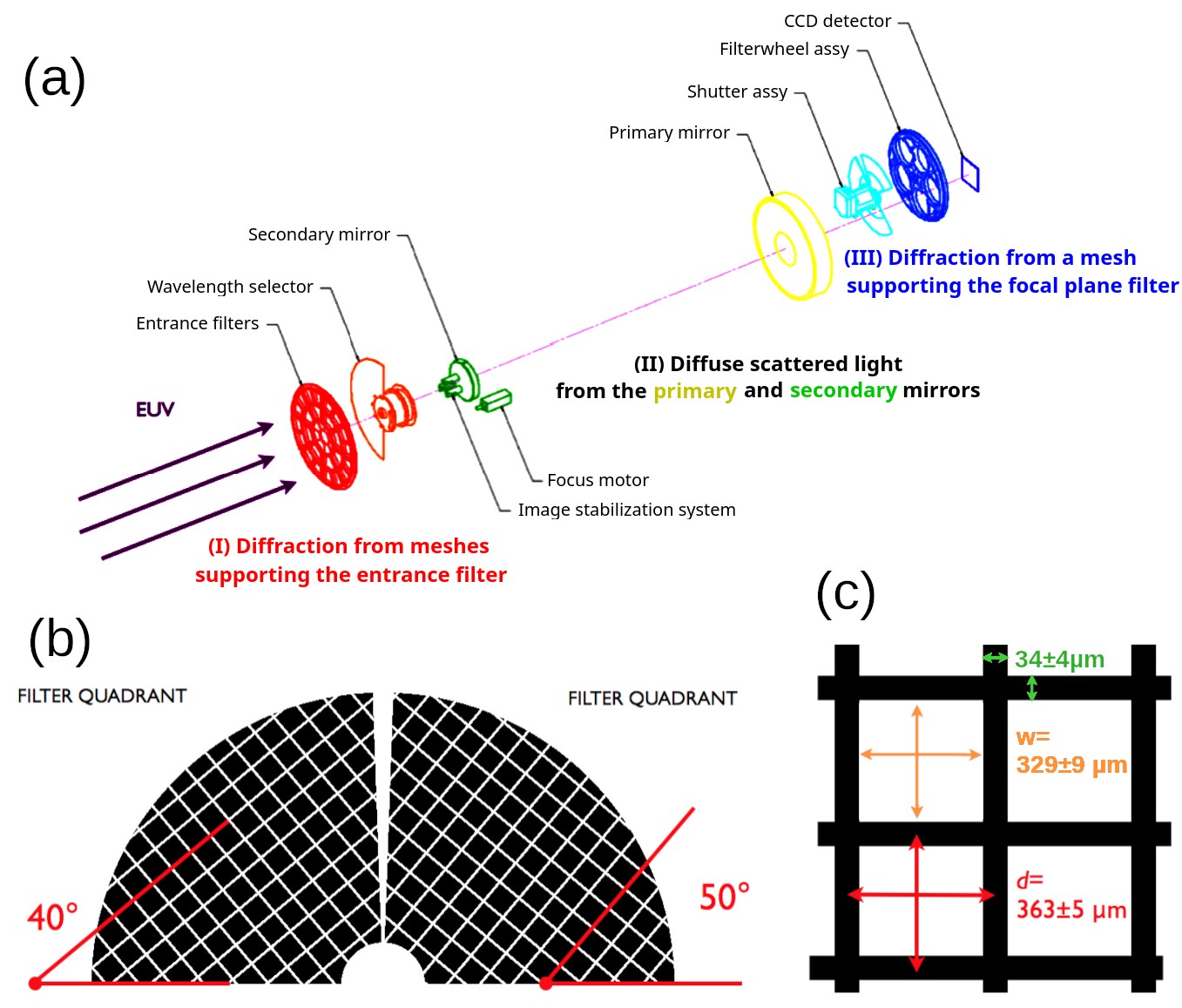} 
    \caption{(a) Schematic of the AIA instrument. The active channel is selected in Telescopes~1,~3, and~4 by the focal plane filter located in the filter wheel and in Telescope~2 by the mechanical wavelength selector behind the entrance filter. Diffraction patterns are caused by two meshes supporting the entrance filter and by the one mesh supporting the focal plane filter. (b): Schematics of the arrangement of the meshes that support the entrance filter. (c) Dimensions of the meshes. Adapted from \citet{grigis2012}}.
    \label{fig:aia}
\end{figure}

AIA consists of four EUV telescopes, each containing multiple observing channels. Telescope~1 observes the~$131$ and \SI{335}{\angstrom}~channels. Telescope~2 observes the~$193$ and \SI{211}{\angstrom}~channels. Telescope~3 observes the \SI{171}{\angstrom} channel, together with the the ultraviolet~$1600$ and~\SI{1700}{\angstrom} and the visible \SI{4500}{\angstrom} channels. These last three channels  are not part of this study. Lastly, Telescope~4 observes the~$94$ and \SI{304}{\angstrom}~channels. 
 
Each telescope uses an on-axis design and has a \SI{20}{cm} diameter primary mirror. However, the optical paths of the individual channels in each telescope are slightly off-axis. This is a result of different reflective coatings having been applied to the top and bottom half of each mirror. Each coating is optimized to reflect the wavelength of one channel and cuts the available aperture per channel in half, resulting in a slight off-axis optical path for each channel.
A schematic of the telescopes is presented in Figure~\ref{fig:aia}. 
Filters are located at the entrance and focal planes of the telescope to absorb visible light and off-band EUV radiation. The entrance filter is supported by two adjacent meshes that are rotated at angles of \SI{40}{\degree} and \SI{50}{\degree} relative to the horizontal direction of the CCD. The focal plane filter in the filter wheel is supported by a single mesh rotated at an angle of \SI{45}{\degree} to this horizontal direction. We define the horizontal direction of the entrance and focal plane meshes as  \SI{40}{\degree}, \SI{50}{\degree}, and \SI{45}{\degree}, respectively, relative to the horizontal CCD direction, and the vertical direction of the meshes as perpendicular to these. 
The number of illuminated wires in each direction in each mesh is about $550$. The center-to-center wire spacing, i.e., the mesh pitch, is \SI{362.9 \pm 5.2}{\um}\footnote{The values in the parentheses give the uncertainties in the last digits of the associated numbers}. The width of the wires is \SI{34.3 \pm 4}{\um}. Accordingly, the width of the mesh window is $\num{328.6 \pm 9.2}$~\si{\um} \citep{grigis2012}.\footnote{\url{https://hesperia.gsfc.nasa.gov/ssw/sdo/aia/idl/psf/DOC/psfreport.pdf}}
The active channel is selected in Telescopes~1,~3, and~4 by a focal plane filter wheel and in Telescope 2 by a mechanical aperture selector which blocks half of the aperture. In Telescope~1 and~4, the focal plane filters do not completely reject light from the other channel in each telescope, resulting in a small amount of cross talk between the channels. The corresponding uncertainty in the measured intensities is typically \SI{<2}{\%} for non-flaring conditions, but can reach up to \SI{40}{\%} for the \SI{335}{\angstrom} channel during flaring conditions \citep{boerner2012}. This cross talk has no significant effect on the calibration of our derived PSFs, as we only use the spacing and intensities of diffraction peaks in flare images for the calibration of the diffraction pattern. The spacing of the diffraction peaks depends approximately linearly on the  wavelength and is thus distinct for each AIA channel on each telescope. Any erroneous diffraction peaks due to cross talk from the complementary channel in each AIA telescope are therefore far outside of the expected calibration parameter space and do not strongly affect our calibration.
The CCD detector of each telescope has a size of $4096\times 4096$ pixels with a plate scale of \SI{0.60}{\arcsec/px} (\SI{2.9e-6}{rad/px}). The CCD is simultaneously read out by 4 read-out amplifiers, where each serves a $2048 \times 2048$ pixels quadrant of the CCD \citep{lemen2012, boerner2012, grigis2012}.

\section{The AIA diffraction pattern} \label{sec:diffraction}

The PSF diffraction pattern of AIA arises from meshes which support the entrance and the focal plane filters. The support of the entrance filter consists of two meshes and the support of the focal plane filter of a single mesh. Therefore, the complete diffraction pattern is given by a combination of three meshes. We first describe the general derivation of the PSF diffraction pattern from the mesh properties, and then calibrate the diffraction pattern of the AIA PSF using solar flare observations.

\subsection{The diffraction pattern} \label{sec:derivingthediffpattern}

\begin{figure}
    \centering
    \includegraphics[width=\textwidth]{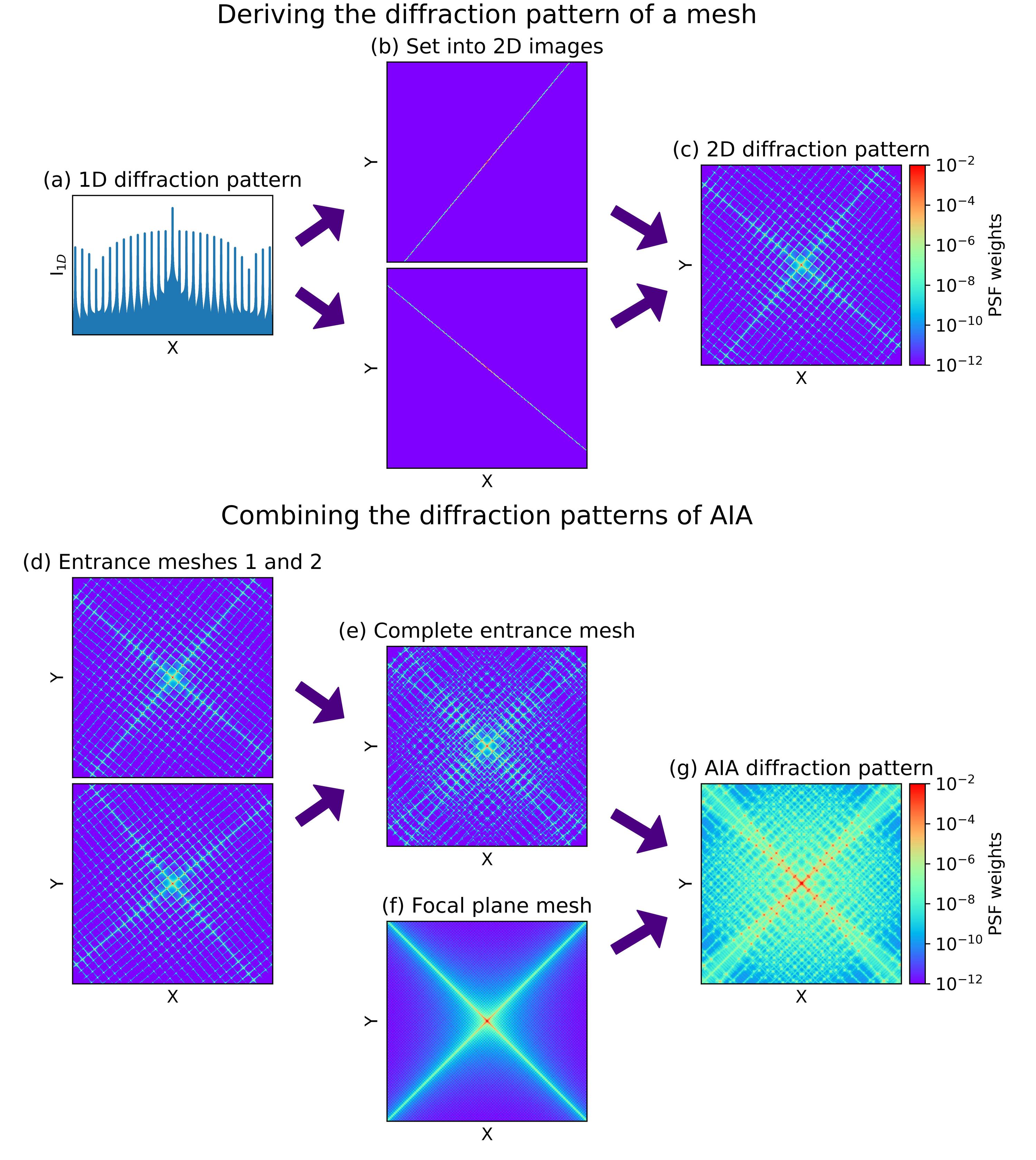}
    \caption{Schematic for the derivation of the AIA PSF diffraction pattern. All images show an enlargement of the region around the PSF center. (a) Derive the 1D diffraction pattern of a grid of periodic slits. (b) Map the 1D diffraction pattern to two 2D images. (c) Convolve the two images to obtain the 2D diffraction pattern of a mesh. (d) Derive the diffraction pattern of the two entrance meshes of AIA. (e) Superimpose these diffraction patterns to obtain the complete diffraction pattern of the entrance mesh. (f) Derive the diffraction pattern of the focal plane mesh. (g) Convolve the diffraction pattern of the entrance mesh with the one from the focal plane mesh to obtain the complete AIA diffraction pattern.}
    \label{fig:diffpattern_derivation}
\end{figure}

We start with the one-dimensional (1D) diffraction pattern of a grid of periodic slits; convolve two perpendicular grids of periodic slits to obtain the two-dimensional (2D) diffraction pattern for a mesh; and finally merge the diffraction patterns of the entrance filter meshes and the focal plane mesh to obtain the complete AIA diffraction pattern (see Figure~\ref{fig:diffpattern_derivation}).
We denote the slit width and mesh window width as $w$; the slit and mesh pitch as $d$; the number of illuminated slits, which is essentially the same as the number of illuminated mesh wires, as $N$; the rotation angle of the mesh to the horizontal CCD direction as $\alpha$; the detector plate scale as $s$; and the wavelength of the light as $\lambda$. 

The 1D diffraction pattern of $N$ periodic slits of finite widths $w$ is given by \citet{bornwolf1999},
\begin{equation}
I_\text{1D} = \sinc^2\left(\frac{w \sin(x)}{ \lambda }\right)\  \left(\frac{\sin(N \pi d \sin(x)/ \lambda)}{\sin(\pi d \sin(x)/ \lambda)} \right)^2, 
\end{equation}
where $I_\text{1D}$ is the observed diffraction pattern intensity on the detector at the 1D location $x$, which runs perpendicular to the orientation of the slits. The diffraction pattern consists of major peaks and minor peaks appearing at 
\begin{align}
x_\text{major} &= n \frac{\arcsin(\lambda / d)}{s}, \label{eq:majpeaks}\\
x_\text{minor} &= \frac{n}{N-1} \frac{\arcsin(\lambda / d)}{s}, \label{eq:minpeaks}
\end{align}
where $n$ is a non-negative integer, the result of $\arcsin(\lambda / d)$ is in radians and $s$ is in radian per pixel, so that the units of $x_\text{major}$ and $x_\text{minor}$ are in pixels. The major and minor peaks both have a full width half maximum $\delta$ of 
\begin{equation}
    \delta \approx \frac{0.885\ \lambda}{N d}\ \frac{1}{s}, \label{eq:widthpeaks}
\end{equation}
where the first term on the right-hand side of Equation~\ref{eq:widthpeaks} is in radians and $s$ is in radians per pixel, so that $\delta$ is in pixels.
To calculate the spacing and the full width half maximum of the diffraction peaks for the seven EUV AIA channels, we use Equations~\ref{eq:majpeaks},~\ref{eq:minpeaks}, and~\ref{eq:widthpeaks}, and assume a grid of periodic slits with the same grid pitch and slit widths as the AIA mesh pitch and mesh window width. 

For the entrance filter, we find a spacing of the major peaks on the CCD of $9$--$32$ px, a spacing of the minor peaks of $0.016$--$0.058$ px, and a full width half maximum of $0.014$--$0.051$ px for both the major and minor peaks. Thus, the major peaks of the entrance filter are well separated on the CCD as shown in Figure~\ref{fig:diffpattern_derivation}(e), but the widths of the minor peaks cannot be resolved when one uses the pixel resolution of AIA. 

The focal plane filter is located in the filter wheel close to the CCD, which results in a magnified diffraction pattern. The magnification factor can be derived from \citet{grigis2012} to be $0.0232$. Considering this magnification factor, the spacing of the focal plane mesh major peaks is $0.2$--$0.8$ px. AIA cannot resolve the location of these major peaks, and thus they appear as a continuous intensity distribution on the CCD, as shown in  Figure~\ref{fig:diffpattern_derivation}(f). 

These estimates show that, when one wants to resolve the diffraction peaks to derive an accurate 1D~diffraction pattern, it has to be derived at a much higher resolution than the AIA pixel resolution. We use a resolution of $\delta/11$~px, which corresponds to $0.001$--$0.04$ px for the entrance filter meshes and $0.00003$--$0.0001$ px for the focal plane filter mesh.

The two-dimensional diffraction pattern of a mesh is given by the convolution of the diffraction patterns of two perpendicular grids of periodic slits \citep{goodman2005}, 
\begin{equation}
    I_\text{2D} = I_\text{1D,$\alpha$} \ast I_\text{1D,$\alpha+\SI{90}{\degree}$},
\end{equation}
where $\ast$ is the convolution operator.
To be able to perform this memory-intensive computation, we bin the derived high-resolution 1D~diffraction pattern to a resolution that is 3~times higher than the AIA CCD, and input it into one array at an angle $\alpha$ and into a second array at an angle $\alpha + \SI{90}{\degree}$. Then, we convolve these two arrays to obtain the 2D~diffraction pattern of a mesh. For the convolution operation, in order to break the periodic boundary conditions in the Fourier domain, the arrays have to be padded with zeros extending beyond each side of the superresolved CCD array by a quarter of their edge lengths. The superresolution of a factor of~3 reduces computational interpolation artifacts during this convolution operation. 

The entrance filter is supported by two adjacent, identical meshes, which are inclined at angles of $\SI{40}{\degree}$ and $\SI{50}{\degree}$  relative to the horizontal CCD direction. The combined diffraction pattern is given by the phase-correct summation of the diffracted light arising from each mesh. However, we do not know with sufficient accuracy the location of the two meshes relative to each other and so are unable able to derive a reasonable wave interference pattern.  Therefore, we neglect the phase component of the diffracted light and approximate the combined diffraction pattern of the two meshes as a simple superposition of the diffraction patterns of the individual meshes. This approximation is justified as the widths of the major diffraction peaks, which contain the majority of the diffracted light, are very small. Their small widths makes it unlikely that major diffraction peaks from the two meshes would overlap and interfere. 
The focal plane filter is supported by a single mesh, which is inclined at an angle of $\SI{45}{\degree}$ relative to the horizontal CCD direction. Convolving the combined diffraction pattern of the entrance meshes with the diffraction pattern of the focal plane mesh at the superresolution of~$3$ and and rebinning the result to the AIA CCD resolution gives the complete diffraction pattern of the AIA meshes. A zoom into the complete diffraction pattern around the  PSF center is shown in Figure~\ref{fig:diffpattern_derivation}(g).

\subsection{Calibrating the AIA diffraction pattern}
The mesh filter parameters given in Section~\ref{sec:aiaoverview} are approximations based on the mechanical drawings. But production and assembling inaccuracies can result in slight deviations. To calibrate the mesh parameters and the associated diffraction patterns, we used images of solar flares where the diffraction pattern were visible. We first describe the composition of the base images for the calibration, then the calibration procedure, and finally the calibration results. A schematic of the calibration procedure is shown in Figure~\ref{fig:overview_diffcal}.

\subsubsection{Base images}

\begin{table}[]
    \centering
    \caption{List of flare images used to calibrate the diffraction pattern.}
    \label{tab:list_of_flares}
        \begin{tabular}{c l c || c l c}
        AIA channel & Flare date & Flare class & AIA channel & Flare date & Flare class  \\ \toprule
        \SI{94}{\angstrom} &  2011 Aug 09 08:13:30 & \multirow{6}{*}{X9.96} &         \SI{94}{\angstrom} &  2015 May 05 22:12:41 & \multirow{7}{*}{X3.93}\\

        \SI{131}{\angstrom} &  2011 Aug 09 08:09:49 & &         \SI{131}{\angstrom} &  2015 May 05 22:11:24 & \\

        \SI{171}{\angstrom} &  2011 Aug 09 08:04:14 & &         \SI{171}{\angstrom} &  2015 May 05 22:10:37 & \\

        \SI{193}{\angstrom} &  2011 Aug 09 08:10:10 & &         \SI{193}{\angstrom} &  2015 May 05 22:10:58 & \\

        \SI{304}{\angstrom} &  2011 Aug 09 08:06:36 & &         \SI{211}{\angstrom} &  2015 May 05 22:11:52 & \\

        \SI{335}{\angstrom} &  2011 Aug 09 08:10:31 & &         \SI{304}{\angstrom} &  2015 May 05 22:09:23 & \\

                            &                      & &         \SI{335}{\angstrom} &  2015 May 05 22:13:18 & \\

        \SI{94}{\angstrom} &  2012 Jan 27 18:47:54 &  \multirow{3}{*}{X2.57} & & \\

        \SI{211}{\angstrom} &  2012 Jan 27 18:50:17 & &         \SI{94}{\angstrom} &  2017 Sep 10 16:09:03 & \multirow{5}{*}{X11.88} \\     
        \SI{335}{\angstrom} &  2012 Jan 27 18:49:43 & & \SI{131}{\angstrom} &  2017 Sep 10 16:07:22 & \\

                            &                      &  & \SI{193}{\angstrom} &  2017 Sep 10 16:02:55 & \\
        \SI{94}{\angstrom} &  2013 May 13 16:06:17 & \multirow{4}{*}{X4.11} &            \SI{211}{\angstrom} &  2017 Sep 10 16:07:26 & \\      

        \SI{131}{\angstrom} &  2013 May 13 16:07:00 & &  \SI{335}{\angstrom} &  2017 Sep 10 16:06:26 & \\

        \SI{171}{\angstrom} &  2013 May 13 16:01:50 & & \\        

        \SI{304}{\angstrom} &  2013 May 13 15:59:47 & &  \SI{94}{\angstrom} &  2023 Mar 03 17:54:39 & \multirow{4}{*}{X2.07}\\ 
        & & & \SI{131}{\angstrom} &  2023 Mar 03 17:54:58 & \\
        \SI{94}{\angstrom} &  2014 Feb 25 00:51:05 & \multirow{5}{*}{X7.13} & \SI{193}{\angstrom} &  2023 Mar 03 17:50:56 & \\
        \SI{131}{\angstrom} &  2014 Feb 25 00:49:24 & & \SI{335}{\angstrom} &  2023 Mar 03 17:57:38 & \\
        \SI{171}{\angstrom} &  2014 Feb 25 00:47:01 & &  \\
        \SI{211}{\angstrom} &  2014 Feb 25 00:53:03 & &  \\
        \SI{304}{\angstrom} &  2014 Feb 25 00:50:11 & & 
        \end{tabular}
\end{table}

\begin{figure}
    \centering
    \includegraphics[width=.72\textwidth]{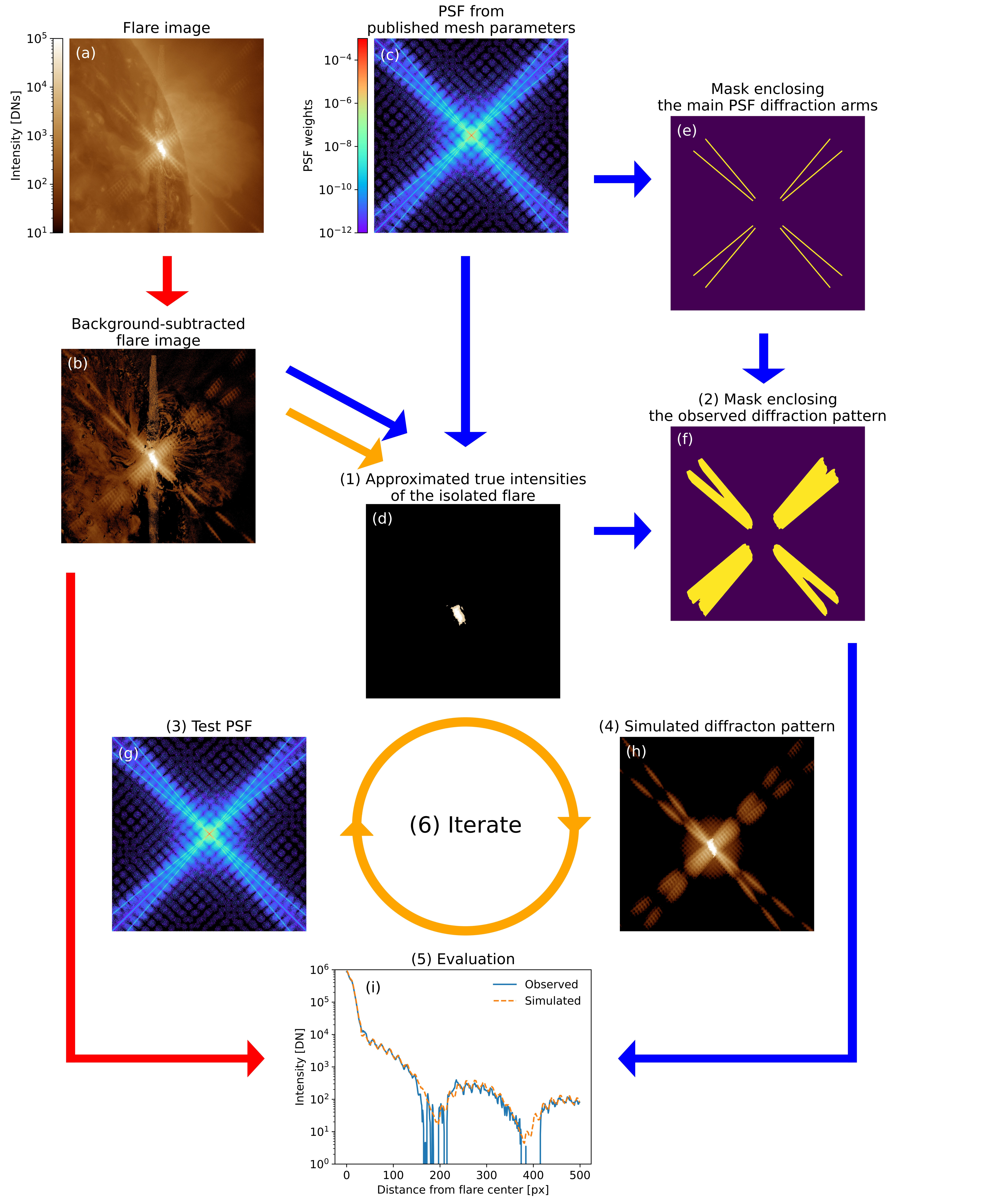}
    \caption{Schematic for the calibration of the mesh parameters. The numbers correspond to steps in the calibration procedure as described in Section~\ref{sec:diffpatterncal}. Red path: We subtract from the solar flare image (a)~a prior non-flare image. This results in (b)~a background-subtracted flare image where the flare and the associated diffraction patterns are clearly visible. This observed diffraction pattern is later used for the evaluation~(i). Blue path: Deconvolving the background-subtracted flare image~(b) with the PSF from the published mesh parameters~(c) and subsequently setting all intensities outside the flare to zero results in an image that contains the approximated true flare intensities of the isolated flare without the diffraction pattern~(d). Convolving a mask whose open area encloses the diffraction arms of the PSF~(e) with this isolated flare image~(d) results in a mask enclosing the approximate location of the observed diffraction pattern~(f). This mask for the observed diffraction pattern will later be used for the evaluation~(i). Orange path: Starting with a test PSF~(g), we first approximate the true flare intensities using this test PSF and isolate the flare~(d). Then, we simulate the flare diffraction pattern by convolving the isolated flare with the test PSF~(h). We evaluate the accuracy of the test PSF by comparing the simulated diffraction pattern with the observed diffraction pattern, using the mask enclosing the observed diffraction pattern~(i). We iterate the procedure of the yellow path with a random walk in the test PSF parameter space to find the mesh parameters that best describe the observed diffraction pattern. }
    \label{fig:overview_diffcal}
\end{figure}

The intensity of the diffraction pattern in solar flare images depends linearly on the intensity of the flare. Only the strongest flares produce clear diffraction patterns that we can use for calibration. Strong flares, on the other hand, saturate the images. Fortunately, during strong flares, AIA takes every second image with a dynamic, short exposure time which can be used to approximate the flare intensities in the saturated images. 

We searched the AIA catalog for the strongest flares that occurred between 2010 and~2023. For each AIA channel, we looked for flare images in which the diffraction pattern was clearly visible compared to the solar background and for which time-adjacent unsaturated short-exposure images were available. An example of such an image is given by the Flare image shown in Figure~\ref{fig:overview_diffcal}(a). We then approximated the unsaturated flare intensities in the full exposure images by a linear interpolation from the short-exposure images. Finally, we reduced the solar background by subtracting the closest-in-time non-flaring image. An example is given by the background-subtracted flare image  shown in Figure~\ref{fig:overview_diffcal}(b). After a final visual check on the quality of the resulting image to verify that the diffraction pattern was clearly visible and the background reduced, we ended up at a dataset of $4$ to $7$ flare images per AIA channel. The dates and strengths of the selected flares are listed in Table~\ref{tab:list_of_flares}.

\subsubsection{Calibration procedure} \label{sec:diffpatterncal}
To calibrate the mesh parameters, we (1) define a function to approximate the true flare intensities and isolate the flare from the surrounding image; (2) create a mask whose open area encompasses the observed diffraction pattern; (3) assume a test set of mesh parameters and derive the associated PSF; (4)  simulate the corresponding flare diffraction pattern; (5) evaluate if the simulated diffraction pattern matches the observed pattern within the mask; and (6) iterate over the steps (3),~(1),~(4),~(5) using a random walk in parameter space to find the best fit. In the following, we describe the process in more detail.

In step~(1), we have developed a routine to approximate the true flare intensities and separate the flare from the diffraction pattern. This process begins by deconvolving the flare image with an input PSF to approximate the true flare intensities. Then, we separate the flare from the observed diffraction pattern by only including those pixels whose intensities are above a threshold level that was adjusted for each flare image individually. The threshold was typically \SI{<20}{\percent} of the flare maximum intensity.  Since the intensity gradient at the flare edge is steep, the exact threshold has only a minor effect on the calibration results.  Afterwards, we set the isolated flare into an empty image. An example for an isolated flare image is shown in Figure~\ref{fig:overview_diffcal}(d).

For step~(2), we derive a mask for the diffraction pattern produced by a given flare. The procedure is outlined by the blue path in Figure~\ref{fig:overview_diffcal}. We begin with constructing a PSF using the published mesh parameters from the instrument team (see Section~\ref{sec:aiaoverview}) and follow the procedure of Section~\ref{sec:derivingthediffpattern}, resulting in Figure~\ref{fig:overview_diffcal}(c). We create a mask for the diffraction arms of the PSF by including only PSF pixels that are more than \SI{100}{px} away from the PSF center but within a distance of $\pm 3$~px to the diffraction arms of the PSF, as shown in Figure~\ref{fig:overview_diffcal}(e). The distance of $100$~px excludes diffraction peaks that are close to the PSF center and thus close to the fast-evolving flare in the observed image. The distance of $3$~px makes the procedure more robust to errors in the published mesh parameters. Then, we derive the isolated flare using the routine from step~(1). Finally, we obtain a mask whose open area encompasses the location of the observed diffraction pattern in the flare image by convolving the isolated flare image with the mask of the PSF diffraction arms. This result is shown in Figure~\ref{fig:overview_diffcal}(f). For each flare, we truncate the length of the diffraction arms in this mask at the point where the observed diffraction pattern is no longer distinguishable from the background. 

In steps~(3)--(5), we simulate the diffraction pattern for a test set of mesh parameters and compare it to the observed diffraction pattern. The procedure is outlined by the orange path in Figure~\ref{fig:overview_diffcal}. We begin by constructing a test PSF from the set of mesh parameters following the procedure of Section~\ref{sec:derivingthediffpattern} and shown in Figure~\ref{fig:overview_diffcal}(g). Then, we approximate the true flare intensities using this PSF and isolate the flare from the diffraction pattern, using the routine defined in step~(1). We obtain a simulated diffraction pattern, shown in Figure~\ref{fig:overview_diffcal}(h), by convolving the isolated flare with the test PSF. Finally, we evaluate the simulated diffraction pattern on the observed diffraction pattern within the mask of the observed diffraction pattern derived in step~(2), as shown in Figure~\ref{fig:overview_diffcal}(i). As an evaluation metric, we use the root mean square percentage error (RMSPE) between the simulated and observed intensities. This RMSPE metric gives all diffraction pattern locations the same weight independent of their intensity and distance from the flaring location.

In step~(6), we fit for the mesh parameters by iterating over the steps (3),~(1),~(4),~(5) for $400$~times using a random walk in the mesh parameter space and minimizing the RMSPE between the simulated and observed diffraction pattern. We chose the random walk method over common regression methods as the latter did not provide stable results, likely due to the diffraction peaks creating a complex solution space with a variety of local minima. For the fitting, we start with the mesh parameters given in Section~\ref{sec:aiaoverview} and a step size of \SI{1}{\um} for the mesh window width, \SI{5}{\um} for the mesh pitch, and \SI{0.2}{\degree} for the mesh angle. Then, we randomly choose a mesh parameter from the set of mesh parameters and modify its value by a random number, which we draw from a Gaussian distribution having a sigma equal the step size. We simulate the flare diffraction pattern from this test set of mesh parameters and evaluated it against the observed intensities. If the RMSPE decreases as compared to the previous iteration, then we accept the test set of mesh parameters as the new start point for the next iteration; if not, then we reject it. For the initial number of steps, equal to 20~times the number of fit parameters, we keep the step size constant. This enables the fit algorithm to sample a large region in the parameter space. Afterwards, we reduce the step size at each iteration exponentially, so that after $400$~iterations in total the fitting has converged to an accuracy \SI{<0.1}{\um}~for the mesh window width, \SI{<0.5}{\um}~for the mesh pitch, and \SI{<0.02}{\degree}~for the mesh angle. 

\subsubsection{Results}

\begin{figure}
    \centering
    \includegraphics[width=\textwidth]{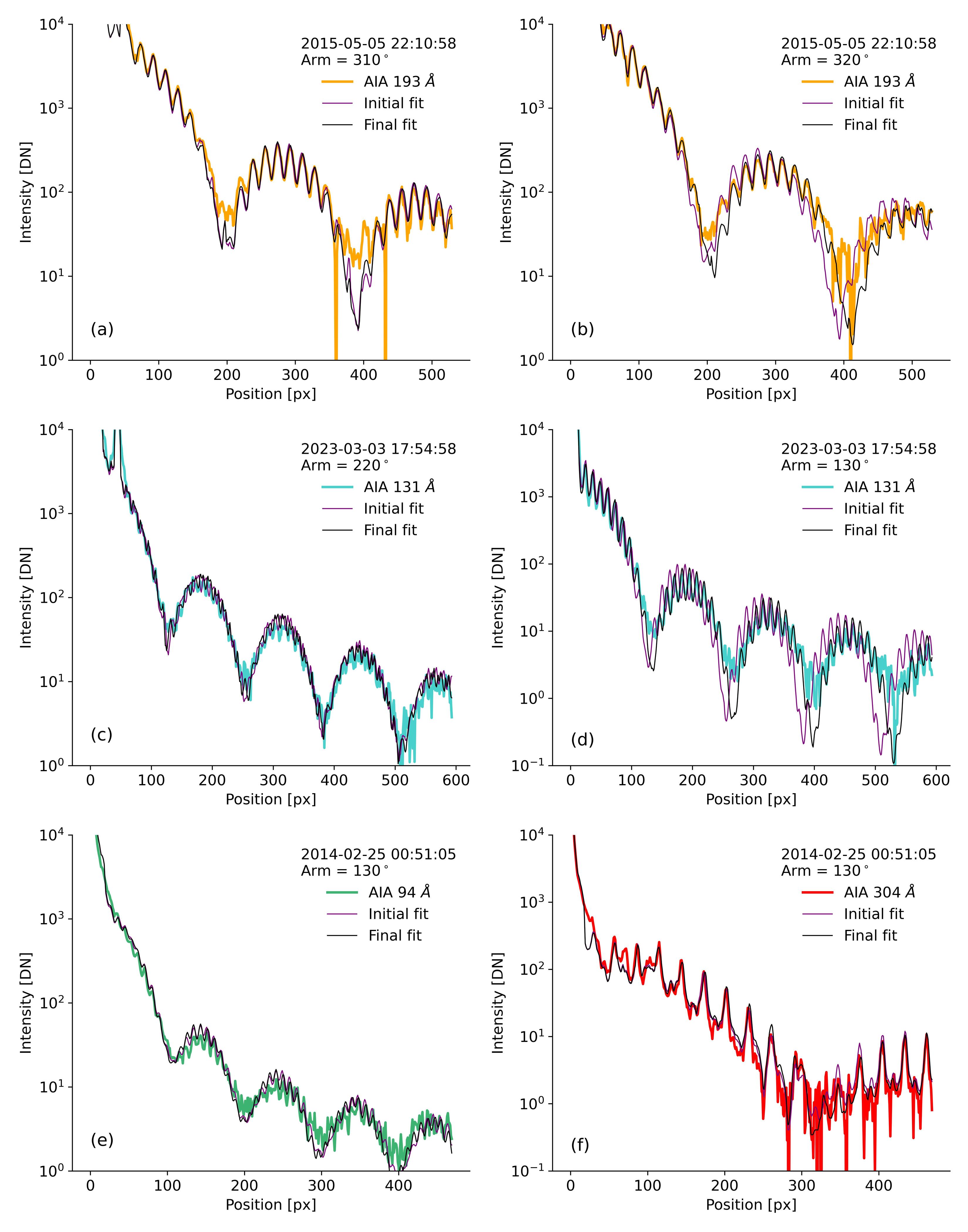}
    \caption{Observed and fitted diffraction pattern intensities for several flares. Flare observed on 2015~May~5 in AIA~\SI{193}{\angstrom}; intensity profiles along the diffraction arms at (a)~\SI{310}{\degree} and (b)~\SI{320}{\degree} are shown. Flare observed on 2023~March~3 in AIA~\SI{131}{\angstrom}; intensity profiles along the diffraction arms at (c)~\SI{220}{\degree} and (d)~\SI{130}{\degree} are shown. Flare observed on 2014~February~25 in AIA (e)~\SI{94}{\angstrom} and (f)~\SI{304}{\angstrom}; intensity profiles along the diffraction arm at \SI{130}{\degree} are shown.}
    \label{fig:fit_diffpattern_problems}
\end{figure}

We have fit for the mesh pitch $d$, which determines the location of the major diffraction peaks; the number of illuminated mesh wires $N$, which determines the number of minor diffraction peaks; the mesh window width $w$, which determines the overall shape of the diffraction orders; and the angle $\alpha$, which determines the direction of the diffraction pattern on the detector. 

Initially, we ran the calibration procedure to fit the mesh parameters from each flare for each AIA channel individually. This assumed that the two entrance meshes and the focal plane mesh in each channel can be described by the same mesh pitch, mesh window width, and number of illuminated mesh wires. From this run, we found that (1) the number of illuminated mesh wires $N$ cannot be determined, as a small change in $N$ has a negligible effect on the PSF diffraction pattern at the resolution of the AIA CCD; (2) the angle of the focal plane filter mesh cannot be precisely determined, as the diffraction pattern from this mesh cannot be resolved by AIA; (3) the fitted mesh parameters vary slightly from flare to flare by more than the uncertainty of the fitting method; (4) the two entrance meshes cannot be described by a single set of mesh parameters; (5) the mesh parameters in the orthogonal directions within each mesh cannot be described by a single set of mesh parameters; and (6) each of the two channels on the same AIA telescope can be described by the same set of mesh parameters.

Points (4)--(6) appeared as a recurrent pattern for all AIA telescopes and all AIA channels. In Figure~\ref{fig:fit_diffpattern_problems}, we illustrate these patterns on the examples of the diffraction patterns from three flares. In Panels~(a) and~(b), we show the observed diffraction patterns from Telescope~2 for the 2015~May~05 flare in the \SI{193}{\angstrom} channel for the diffraction arm at \SI{310}{\degree}, which belongs to Entrance Mesh~1, and for the diffraction arm at \SI{320}{\degree}, which belongs to Entrance Mesh~2. The predicted diffraction patterns from our initial fits, which are shown in Figure~\ref{fig:fit_diffpattern_problems} by the purple lines, assumed the same mesh pitches and mesh window widths for both entrance meshes. For Entrance Mesh~1, the predicted diffraction pattern fits well to the observed diffraction pattern for the location of the diffraction peaks and for the overall shape of the intensity, i.e., the shape of the diffraction orders. However, for Entrance Mesh~2, the predicted diffraction pattern slightly deviates from the observed shape. This deviation indicates that Mesh~2 has a slightly smaller average mesh window width than Mesh~1.  Both, however, are well within the tolerance given by the manufacturer.

In Panels~(c) and~(d), we show the observed diffraction patterns from Telescope~1 for the 2023~March~3 flare in the \SI{131}{\angstrom} channel for both the diffraction arm at \SI{220}{\degree} and the diffraction arm at \SI{130}{\degree}, which are associated with the vertical and horizontal mesh directions within Entrance Mesh~1, respectively. For the diffraction arm at \SI{220}{\degree}, the overall shape of the predicted diffraction pattern for the initial fits resembles reasonably well the observed diffraction pattern. However, for the diffraction arm at \SI{130}{\degree}, the predicted shape clearly differs from the observed shape, indicating that the horizontal mesh direction has a smaller average mesh window width than the vertical direction. These results likely indicate a slight inhomogeneity during the production process of the mesh wires, which is, however, well  within the tolerance given by the manufacturer.

In Panels~(e) and~(f), we show the observed diffraction pattern from Telescope~4 for the 2014~February~25 flare along the diffraction arm at \SI{130}{\degree} for the~$94$ and \SI{304}{\angstrom} channels, respectively. Both these channels are located on Telescope~4 and thus share the same entrance meshes. We fitted the mesh parameters for the \SI{94}{\angstrom} channel and predicted from these the diffraction pattern of the \SI{304}{\angstrom} channel. The predictions reproduces well the observed diffraction patterns regarding the location of the diffraction peaks and the shape of the diffraction orders. This suggests that we can fit both channels of Telescope~4 with the same mesh parameters.

\begin{table}[]
    \centering
    \caption{Fitted mesh parameters}
    \label{tab:fitted_mesh_parameters}
    \begin{tabularx}{.9\textwidth}{c c | c c c c}
       \multicolumn{2}{c | }{\multirow{2}{*}{Parameter}} &  Telescope 1 & Telescope 2 & Telescope 3 & Telescope 4 \\
       &  &  \SI{131}{\angstrom}, \SI{335}{\angstrom} & \SI{193}{\angstrom}, \SI{211}{\angstrom} & \SI{171}{\angstrom} &
        \SI{94}{\angstrom}, \SI{304}{\angstrom}\\ \toprule

        \multicolumn{2}{c | }{Entrance Mesh 1} & & \\[4pt]
        \multirow{3}{*}{\quad \quad horizontal} & $\alpha$ & \SI{39.65 \pm 0.01}{\degree} & \SI{40.12 \pm 0.03}{\degree} & \SI{40.02 \pm 0.07}{\degree} & \SI{40.19 \pm 0.03}{\degree} \\
        & $d$ & \SI{362.7 \pm 0.2}{\um} & \SI{362.3 \pm 0.2}{\um} & \SI{362.0 \pm 0.3}{\um} & \SI{362.5 \pm 0.4}{\um} \\
        & $w$ & \SI{329.3 \pm 0.1}{\um} & \SI{328.2 \pm 0.1}{\um} & \SI{328.6 \pm 0.3}{\um} & \SI{329.9 \pm 0.2}{\um} \\[4pt]

        \multirow{3}{*}{\quad \quad vertical} & $\alpha$ & \SI{129.65 \pm 0.03}{\degree} & \SI{130.11 \pm 0.03}{\degree} & \SI{130.05 \pm 0.05}{\degree} & \SI{130.12 \pm 0.03}{\degree} \\
        & $d$ & \SI{362.5 \pm 0.2}{\um} & \SI{362.8 \pm 0.4}{\um} & \SI{362.4 \pm 0.8}{\um} & \SI{362.4 \pm 0.4}{\um} \\
        & $w$ & \SI{327.7 \pm 0.1}{\um} & \SI{328.1 \pm 0.2}{\um}& \SI{329.6 \pm 0.2}{\um} & \SI{331.0 \pm 0.1}{\um} \\[8pt]

        \multicolumn{2}{c | }{Entrance Mesh 2} & & \\[4pt]
        \multirow{3}{*}{\quad \quad horizontal} & $\alpha$ & \SI{49.97 \pm 0.03}{\degree} & \SI{50.39 \pm 0.09}{\degree} & \SI{50.33 \pm 0.14}{\degree} & \SI{50.07 \pm 0.04}{\degree} \\
        & $d$ & \SI{362.5 \pm 0.2}{\um} & \SI{362.6 \pm 0.6}{\um} & \SI{360.7 \pm 0.5}{\um} & \SI{362.7 \pm 0.2}{\um}\\
        & $w$ & \SI{331.5 \pm 0.1}{\um} & \SI{330.2 \pm 0.1}{\um} & \SI{328.2 \pm 0.2}{\um} & \SI{330.9 \pm 0.2}{\um} \\[4pt]

        \multirow{3}{*}{\quad \quad vertical} & $\alpha$ & \SI{140.00 \pm 0.02}{\degree} & \SI{140.35 \pm 0.08}{\degree} & \SI{140.23 \pm 0.05}{\degree} & \SI{139.93 \pm 0.04}{\degree} \\
        & $d$ & \SI{362.4 \pm 0.3}{\um} & \SI{362.7 \pm 0.4}{\um} & \SI{362.1 \pm 0.9}{\um} & \SI{362.2 \pm 0.4}{\um} \\
        & $w$ & \SI{330.0 \pm 0.1}{\um} & \SI{329.0 \pm 0.3}{\um} & \SI{329.2 \pm 0.5}{\um} & \SI{329.4 \pm 0.2}{\um}\\[8pt]
        \multicolumn{2}{c | }{Diffracted light} & \SI{27.19 \pm 0.01}{\percent}, \SI{33.24 \pm 0.02}{\percent} & \SI{30.33 \pm 0.04}{\percent}, \SI{30.40 \pm 0.04}{\percent} & \SI{29.96 \pm 0.09}{\percent} & \SI{24.34 \pm 0.02}{\percent}, \SI{30.08 \pm 0.03}{\percent}
    \end{tabularx}
\end{table}

Following these results, we adjusted the calibration procedure. We fixed the number of illuminated mesh wires and the mesh parameters of the focal plane mesh to their assumed values from the mechanical drawings given in  Section~\ref{sec:aiaoverview}. Then, we fit for the mesh window width, the mesh pitch, and the mesh angle for each entrance mesh and for each direction within the entrance meshes. This results in a total of 12~fit parameters. We simultaneously fit for these 12 mesh parameters from all flare images of both channels within a given AIA telescope, which increases the statistical significance. In the longer-wavelength channels, the spacing of the diffraction peaks are larger, which allows the mesh pitches to be better resolved. The shorter-wavelength channels contain more diffraction orders, so that the mesh window widths can be more precisely determined. We estimated the uncertainty in the fit parameters by repeating the fit by bootstrapping the input flare images, i.e., by resampling the dataset of the input flare images with replacement, for a total of 10 iterations. 

For each AIA channel, the fit results and the amount of light that is diffracted by the meshes are given in Table~\ref{tab:fitted_mesh_parameters}. The diffracted light is about $24$~to~\SI{33}{\percent} with an uncertainty of \SI{<0.1}{\percent}.
The new predicted diffraction patterns from these fits are shown  for the previous examples in Figure~\ref{fig:fit_diffpattern_problems} by the black lines, which now represent robust fits to the observed diffraction patterns. 
We note that these fit results represent the values and uncertainties for the mesh parameters as fitted from all flares simultaneously under study, which make the fit more robust. However, we surmise that the mesh parameters may change slightly over time due to mechanical and temperature deformations. For any individual flare, the associated mesh parameters might deviate beyond the given errors. This will affect the exact location of the observed diffraction pattern, but has a negligible effect on the amount of light that is diffracted by the meshes.

\section{The Diffuse Scattered Light in AIA} \label{sec:diffuse}

Diffuse scattered light within the AIA instrument is assumed to be caused by the scattering of photons on the micro-roughness of the mirrors. Since the wavelength of the AIA channels is the same order of magnitude as the micro-roughness, this scattering is expected to be strong, resulting in a large amount of diffuse scattered light, which is scattered over the entire detector.

The effect of diffuse scattered light is clearly seen in AIA images that are partially occulted, such as the lunar transits shown in Figure~\ref{fig:tail_problems}. The intensities within the lunar-occulted pixels should be zero. Thus, any residual intensities measured there are associated with either errors in the instrumental calibration or with diffuse scattered light from the exposed portion of the image into the lunar-occulted portion. 

In the following subsections, we use the residual intensities in the lunar-occulted portion of the image to reconstruct the function describing the diffuse scattered light. We first discuss issues in the instrumental calibration; then describe the inversion algorithm; and finally present the results.

\subsection{Issues in the instrumental calibration} \label{sec:instrumental_issues}

\begin{figure}
    \centering
    \includegraphics[width=\textwidth]{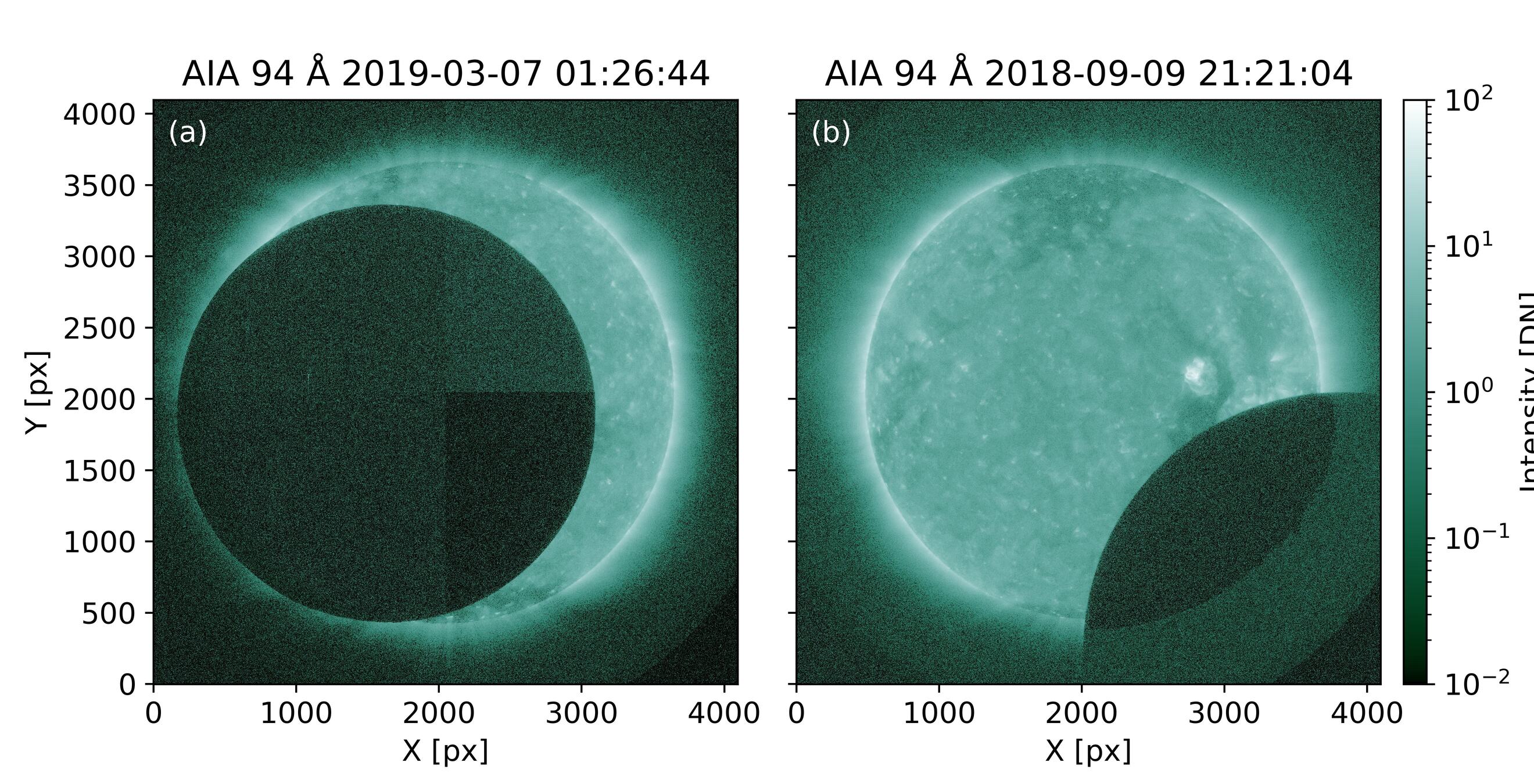}
    \caption{Issues in the calibration of AIA. (a) Different residual offsets in the four quadrants of the AIA image. (b) Ghost image: the solar limb is visible in the occulted region. In addition, the corners of the image are affected by vignetting.}
    \label{fig:tail_problems}
\end{figure}

To fit the diffuse scattered light, we evaluate the residual intensities measured in lunar-occulted pixels. These intensities are very small and typically at the noise level. At these low intensities, residual errors in the instrumental calibration matter. 

We have encountered 3 issues in the instrumental calibration: (1) different intensity offsets in the four quadrants of the AIA images; (2) ghost images due to outdated calibrations; and (3) vignetting in the image corners:
\begin{itemize}
    \item Different intensity offsets in the four image quadrants result from small errors in the calibrations of the four read-out amplifiers, i.e., one for each CCD quadrant. These offsets are at the order of $0.2$~digital numbers (DNs), but can be seen in very-low intensity regions such as lunar occultations, as shown in Figure~\ref{fig:tail_problems}(a). 
    \item Ghost images in AIA are a result of outdated dark current and flat field calibrations. AIA performs dark current and flat field calibrations about once every 3\ --\ 4\ months. However, the CCD behavior evolves over that time, diverging from the most recent calibration data. This small calibration error is typically smaller than \SI{1}{\percent} of the solar disk intensity. It becomes visible as a faint ghost image that still shows the Sun in the lunar-occulted portion of the image, as shown in Figure~\ref{fig:tail_problems}(b). 
    \item Vignetting of the image corners, as shown in Figure~\ref{fig:tail_problems}(a) and~(b), are caused by shading from the filter wheel mechanism \citep{lemen2012}. As such, the measured intensities in the image corners are not reliable and should be excluded from any analysis.
\end{itemize}

The artifacts introduced by these issues are small, but they affect the measured intensity distribution in the low-intensity lunar-occulted portion of the images and thereby distort the reconstruction results for the portion of the PSF that describes the diffuse scattered light. To mitigate the effects of the artifacts, we only use lunar transit images where the total photon count is large, i.e., where artifacts have a comparably small effect. In addition, we fit for the diffuse scattered light from several lunar transit images simultaneously, which further decreases the effect of calibration errors.

\subsection{The inversion algorithm}

We use the inversion algorithm of \citet{hofmeister2022} to fit the portion of the PSF that describes the diffuse scattered light. The inversion is based on a variation of the PSF defining equation, 
\begin{align}
I_\text{o,$\pmb{r}$} &= \sum_{S}  \widetilde{\text{psf}}_S \sum_{\pmb{\Delta r} \text{ in } S} I_\text{t,$\pmb{r}$-$\pmb{\Delta r}$}  + \left( 1 - \sum_S n_S\ \widetilde{\text{psf}}_S\right) \sum_{\pmb{\Delta r}}  I_\text{t,$\pmb{r}$-$\pmb{\Delta r}$} \ \overline{\text{psf}}_{\pmb{\Delta r}} + \epsilon \\ &= \sum_{S}  \widetilde{\text{psf}}_S\ A_S  + \left( 1 - \sum_S n_S\ \widetilde{\text{psf}}_S\right) B. \label{eq:psf_px3}
\end{align}
Here, $I_\text{o,$\pmb{r}$}$~denotes the observed intensity in a pixel at a location $\pmb{r}$; $I_\text{t}$~denotes the true intensity in a pixel, i.e., the intensity that would have been measured without instrumental effects; $\widetilde{\text{psf}}_S$~are the PSF weights to fit;  $\overline{\text{psf}}$~is a known PSF contribution, such as the diffraction pattern; $S$~is a spatial segment of the PSF;  $n_S$~is the number of pixels in a PSF segment; $\pmb{\Delta r}$~is a distance vector in the PSF as measured from the PSF center; and $\epsilon$~is the noise component. In the second line of Equation~\ref{eq:psf_px3}, we have merged the known terms, i.e., the terms that can be derived from the image, into the coefficients $A_S$ and $B$.

Equation~\ref{eq:psf_px3} holds for each pixel in the image plane and thus set up systems of equations. Within these systems of equations, the observed intensities are know in the entire image and the true intensities are known to be zero within the occulted pixels. The true intensities in the unocculted portion of the image can be approximated by a semi-blind iterative procedure, where one alternately fits the PSF from an approximated true image and afterwards uses the fitted PSF to update the approximated true image. Employing the a priori knowledge that the true intensity within the occulted region is zero makes this procedure robust \citep{hofmeister2022}. The PSF coefficients $\widetilde{\text{psf}}_S$ can be determined by a multi-linear fit to the system of equations, provided that we choose a reasonable PSF segmentation, that we have fewer PSF segments than occulted pixels, and that we can reduce the noise component $\epsilon$ in the equation to a negligible level.

\begin{figure}
    \centering
    \includegraphics[width=\textwidth]{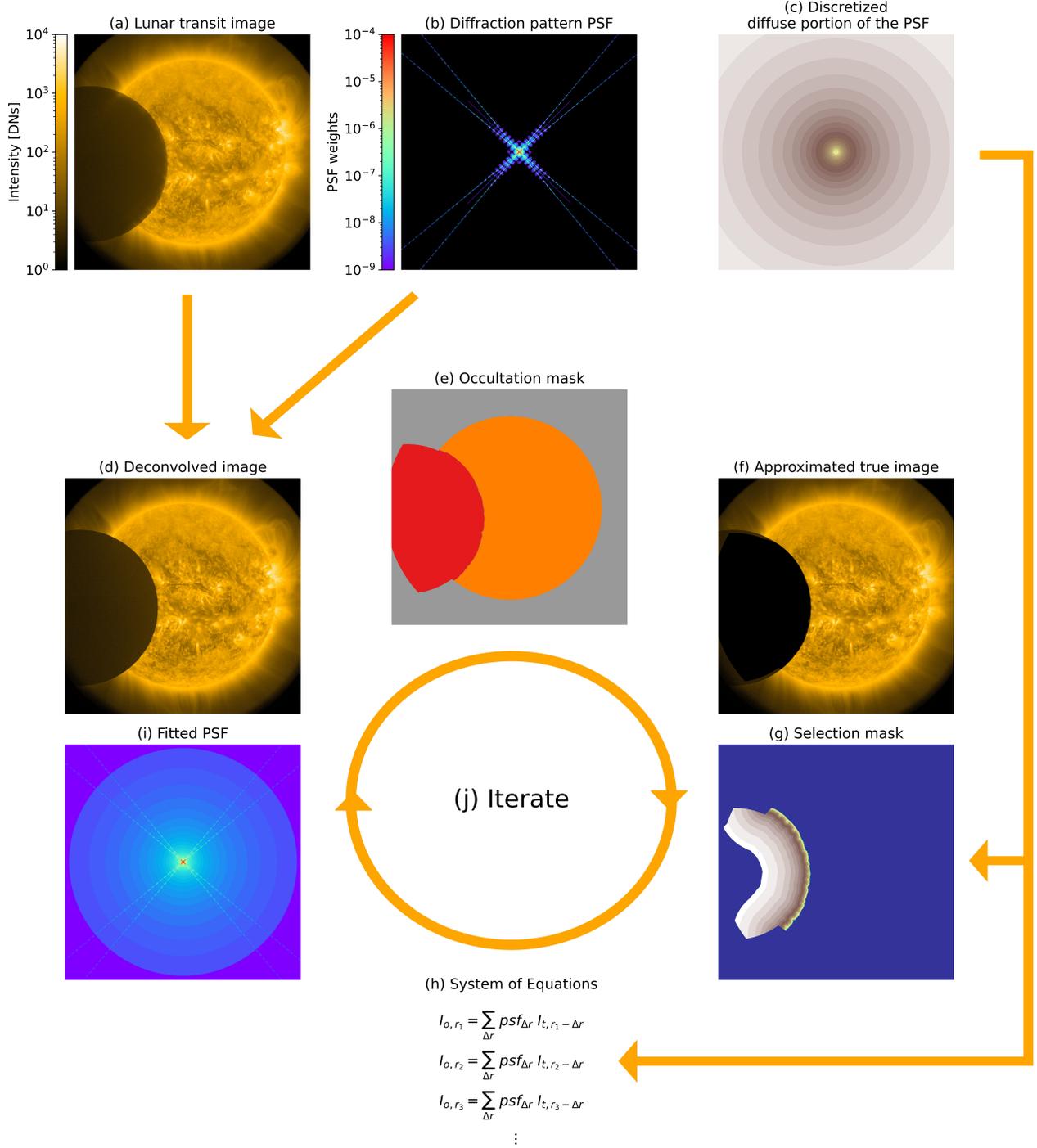}
    \caption{Flow diagram for the calibration procedure of the diffuse scattered light portion of the PSF. The input includes (a)~a~lunar transit image, (b)~the PSF describing the diffraction pattern of the meshes, and (c)~a discretization of the diffuse portion of the PSF, where each segment corresponds to one PSF coefficient to fit. We first derive a PSF deconvolved image~(d) and a mask containing the lunar-occulted and the solar on-disk pixels~(e). By setting the occulted pixels in the deconvolved image to zero, we obtain an approximation of the true image~(f). We determine the location of specific occulted pixels which we will use to fit the PSF, and call this mask the Selection Mask~(g). Using the PSF discretization from (c), the approximated true image from (f), and the Selection Mask from (g), we set up a system of equations which describes the scattered light~(h). By solving this system of equations, we obtain an approximation for the diffuse scattered portion of the PSF. Combining the diffuse scattered portion with the PSF diffraction pattern, we obtain a new approximation for the PSF~(i). We use this PSF to derive a better approximation for the true image, and iterate this procedure until the fit has converged~(j).}
    \label{fig:tail_overview}
\end{figure}

To derive the PSF coefficients from the observed intensity distributions within lunar occultations, we have to perform the following steps: (a) select appropriate lunar transit images, (b) calculate and fix the  diffraction pattern portion of the PSF, (c) discretize the to-be-fitted diffuse scattered light portion of the PSF spatially into segments, (d) deconvolve the lunar transit images with the PSF, (e) identify the occulted pixels in the images, (f) approximate the true images, (g) select a subset of the occulted pixels for setting up the system of equations, (h) set up the system of equations, (i) fit the PSF from the system of equations, and (j) repeat steps (d)\ --\ (i) until the solution for the approximated true image and the fitted PSF have converged. In Figure~\ref{fig:tail_overview}, we visualize these steps by showing a schematic of the procedure; the label in each panel matches the corresponding step in our procedure.

In the following paragraphs, we explain each of these steps in more detail. 

\paragraph{(a) Select appropriate lunar transit images}
To minimize the effects of calibration errors and noise, we only employ lunar transit images with comparably large intensities in the occulted pixels. The measured intensity in the occulted pixels due to scattered light depends on 3 factors: (1)~the intensities in the unocculted portion of the image, which is the source of the scattering and which depends on the solar cycle, (2)~the sensitivity of the instrument, which degrades over time, and (3)~the image exposure time. 
Following these considerations, we only selected lunar transit images from 2010 to~2014, i.e., in the first years after the launch of SDO up to solar maximum, guaranteeing a good sensitivity of the instrument with large intensities in the unocculted pixels.  

 In this time range, we found 5 lunar transits during which AIA took long observations with exposure times of $3-7$~s. These images have the strongest signal by far in the lunar-occulted region from all available AIA images, which enables us to accurately fit the long-distance scattered portion of the diffuse light, i.e., the PSF tail.  As the moon covers different CCD quadrants during its transit and since each quadrant has its own calibration errors, we decided to select up to two AIA images per lunar transit where the moon occulted different portions of the CCD. This resulted in a total of 8\ -- 10\ long-exposure lunar transit images per AIA channel (Table~\ref{tab:list_of_transits_longexp}). 
 
 In addition, we selected 5~short-exposure lunar transit images with exposure times of $0.5-2$~s (Table~\ref{tab:list_of_transits_shortexp}). In these images, the moon moved by less than \SI{1}{px}, which reduces the blurring of the lunar-solar boundary. We use these images to fit for the short- to medium-distance scattered portion of the diffuse scattered light, i.e., the PSF wing.

\begin{table}[]
    \centering
    \caption{List of long-exposure lunar transit images used to calibrate the diffuse scattered light portion of the PSF.}
    \label{tab:list_of_transits_longexp}
        \begin{tabular}{c l c || c l c}
        AIA channel & Lunar transit date & Exposure time &  AIA channel & Lunar transit date & Exposure time  \\ \toprule
\multirow{10}{*}{\SI{94}{\angstrom}} & 2010 Oct 07 11:44:47 & \multirow{10}{*}{\SI{7}{s}} &	\multirow{10}{*}{\SI{211}{\angstrom}}	& 2010 Nov 06 06:47:16 &	\multirow{10}{*}{\SI{5}{s}} \\
 & 2010 Nov 06 06:45:47 & & & 2012 Feb 21 13:46:36 & \\
 & 2010 Nov 06 06:48:07 & & & 2012 Feb 21 14:20:36 & \\
 & 2011 May 03 07:19:47 & & & 2012 Apr 21 07:38:56 & \\
 & 2011 May 03 07:35:47 & & & 2012 Apr 21 07:32:36 & \\
 & 2012 Feb 21 14:31:27 & & & 2013 Mar 11 11:49:35 & \\ 
 & 2013 Mar 11 11:48:06 & & & 2013 Mar 11 12:07:35 & \\
 & 2013 Mar 11 12:07:26 & & & 2014 Sep 24 07:09:35 & \\
 & 2014 Sep 24 06:57:06 & & & 2014 Nov 22 22:43:15 & \\ 
 & 2014 Sep 24 07:15:46 & & & 2014 Nov 22 22:49:55 & \\ 
 &                     & & &                     & \\ 

\multirow{9}{*}{\SI{131}{\angstrom}} & 2010 Oct 07 11:41:20 & \multirow{10}{*}{\SI{7}{s}} &	\multirow{10}{*}{\SI{304}{\angstrom}}	& 2010 Oct 07 11:34:59 &	\SI{5}{s} \\ 
 & 2010 Nov 06 06:49:40 & & & 2010 Oct 07 11:41:19 & \SI{5}{s}\\
 & 2010 Oct 07 11:51:00 & & & 2010 Nov 06 06:49:19 & \SI{5}{s}\\
 & 2011 May 03 07:20:00 & & & 2010 Nov 06 06:51:59 & \SI{5}{s}\\
 & 2011 May 03 07:37:20 & & & 2011 May 03 07:19:59 & \SI{5}{s}\\
 & 2013 Mar 11 11:57:19 & & & 2011 May 03 07:29:39 & \SI{5}{s}\\
 & 2013 Mar 11 12:19:59 & & & 2013 Mar 11 12:03:36 & \SI{7}{s}\\
 & 2014 Sep 24 07:01:19 & & & 2013 Mar 11 12:13:56 & \SI{7}{s} \\
 & 2014 Sep 24 07:10:59 & & & 2014 Sep 24 06:58:56 & \SI{7}{s}\\
 &                     & & & 2014 Sep 24 07:04:56 & \SI{7}{s}\\
\multirow{10}{*}{\SI{171}{\angstrom}} &  2010 Nov 06 06:55:15 &	\multirow{10}{*}{\SI{3}{s}} & & & \\
 & 2010 Oct 07 11:26:35 & & \multirow{8}{*}{\SI{335}{\angstrom}} & 2010 Oct 07 11:32:50 & \multirow{8}{*}{\SI{7}{s}} \\
 & 2011 May 03 07:21:35 & & & 2010 Oct 07 12:00:50 & \\
 & 2011 May 03 07:33:55 & & & 2010 Nov 06 06:53:10 & \\
 & 2012 Feb 21 14:02:55 & & & 2011 May 03 07:35:10 & \\
 & 2012 Apr 21 07:32:35 & & & 2011 May 03 07:37:30 & \\
 & 2013 Mar 11 11:49:34 & & & 2013 Mar 11 11:55:49 & \\
 & 2013 Mar 11 11:59:54 & & & 2013 Mar 11 12:17:49& \\
 & 2014 Sep 24 06:57:34 & & & 2014 Sep 24 07:04:49 & \\
 & 2014 Sep 24 07:13:14 & & & & \\
 &                     & & & & \\
\multirow{10}{*}{\SI{193}{\angstrom}} &  2010 Oct 07 11:29:28 &	\multirow{10}{*}{\SI{3}{s}} & & & \\ 
 & 2010 Oct 07 12:09:08 & & & & \\
 & 2010 Nov 06 06:47:48 & & & & \\			
 & 2010 Nov 06 06:56:08 & & & & \\			
 & 2011 May 03 07:23:28 & & & & \\
 & 2011 May 03 07:37:48 & & & & \\
 & 2012 Feb 21 13:45:08 & & & & \\
 & 2012 Feb 21 14:14:48 & & & & \\
 & 2012 Apr 21 07:32:48 & & & & \\
 & 2013 Mar 11 12:21:47 & & & & 
        \end{tabular}
\end{table}

\begin{table}[]
    \centering
    \caption{List of short-exposure lunar transit images used to calibrate the diffuse scattered light portion of the PSF.}
    \label{tab:list_of_transits_shortexp}
        \begin{tabular}{c l c || c l c}
        AIA channel & Lunar transit date & Exposure time &  AIA channel & Lunar transit date & Exposure time  \\ \toprule
\multirow{10}{*}{\SI{94}{\angstrom}} &  2010 Nov 06 06:45:10 & \multirow{10}{*}{\SI{2}{s}} &	\multirow{10}{*}{\SI{211}{\angstrom}}	& 2010 Oct 07 11:29:39 &	\multirow{10}{*}{\SI{1}{s}} \\
 & 2010-12-06 03:08:10 & & & 2010 Oct 07 12:06:19 & \\
 & 2011 May 03 07:19:10 & & & 2010 Nov 06 06:52:59 & \\
 & 2011 May 03 07:32:30 & & & 2011 May 03 07:30:39 & \\
 & 2012 Feb 21 14:48:10 & & & 2012 Feb 21 13:54:19 & \\
 & 2012 Apr 21 07:36:10 & & & 2012 Feb 21 14:14:19 & \\ 
 & 2013 Mar 11 11:46:29 & & & 2012 Apr 21 07:35:59 & \\
 & 2013 Mar 11 12:12:29 & & & 2012 Apr 21 07:36:19 & \\
 & 2014 Sep 24 06:56:09 & & & 2013 Mar 11 12:24:38 & \\ 
 & 2014 Sep 24 07:16:09 & & & 2014 Sep 24 07:11:58 & \\ 
 &                     & & &                     & \\ 

\multirow{9}{*}{\SI{131}{\angstrom}} & 2010 Oct 07 11:31:42 & \multirow{10}{*}{\SI{2}{s}} &	\multirow{10}{*}{\SI{304}{\angstrom}}	& 2010 Oct 07 11:30:21 & \multirow{10}{*}{\SI{1}{s}} \\ 
 & 2011 May 03 07:23:02 & & & 2010 Oct 07 11:52:01 & \\
 & 2011 May 03 07:29:02 & & & 2010 Nov 06 06:50:21 & \\
 & 2012 Apr 21 07:31:42 & & & 2010 Nov 06 06:52:21 & \\
 & 2012 Apr 21 07:38:22 & & & 2011 May 03 07:25:01 & \\
 & 2013 Mar 11 12:07:01 & & & 2011 May 03 07:30:41 & \\
 & 2013 Mar 11 12:14:41 & & & 2013 Mar 11 12:02:20 & \\
 & 2014 Sep 24 07:02:21 & & & 2013 Mar 11 12:16:20 &  \\
 & 2014 Sep 24 07:12:21 & & & 2014 Sep 24 07:06:40 & \\
 & 2014 Nov 22 22:55:01 & & & 2014 Sep 24 07:10:00 & \\
 &                     & & &                     & \\ 

\multirow{10}{*}{\SI{171}{\angstrom}} &  2010 Oct 07 11:28:17 &	\multirow{10}{*}{\SI{0.5}{s}} &  \multirow{8}{*}{\SI{335}{\angstrom}} & 2010 Oct 07 11:42:12 & \multirow{8}{*}{\SI{2}{s}}\\
 & 2010 Oct 07 12:05:37 & & & 2010 Oct 07 11:50:12 & \\
 & 2010 Nov 06T 06:54:57 & & & 2010 Nov 06 06:36:12 & \\
 & 2011 May 03 07:22:17 & & & 2010 Nov 06 06:42:52 & \\
 & 2011 May 03 07:36:57 & & & 2011 May 03 07:18:12 & \\
 & 2013 Mar 11 11:54:56 & & & 2011 May 03 07:36:32 & \\
 & 2013 Mar 11 12:22:56 & & & 2012 Feb 21 13:10:12 & \\
 & 2014 Jan 30 15:21:49 & & & 2012 Feb 21 14:45:32 & \\
 & 2014 Jan 30 15:22:37 & & & 2013 Mar 11 11:42:31 & \\
 & 2014 Sep 24 07:01:56 & & & 2013 Mar 11 12:36:31 & \\
 &                     & & & & \\
\multirow{10}{*}{\SI{193}{\angstrom}} &  2010 Oct 07 11:56:29 &	\multirow{10}{*}{\SI{0.5}{s}} & & & \\ 
 & 2010 Nov 06 06:47:09 & & & & \\
 & 2010 Nov 06 06:54:49 & & & & \\			
 & 2011 May 03 07:32:49 & & & & \\			
 & 2012 Feb 21 13:50:29 & & & & \\
 & 2012 Apr 21 07:38:09 & & & & \\
 & 2012 Feb 21 14:14:09 & & & & \\
 & 2013 Mar 11 11:50:28 & & & & \\
 & 2013 Mar 11 12:08:48 & & & & \\
 & 2014 Sep 24 06:58:48 & & & & 
        \end{tabular}
\end{table}

\paragraph{(b) Calculate the known diffraction pattern PSF}
Since the diffraction pattern of AIA is accurately known, we derive the diffraction pattern following the procedure of Section~\ref{sec:diffraction} and use it as the known PSF contribution in Equation~\ref{eq:psf_px3}. 

\paragraph{(c) Discretize the to-be-fitted diffuse portion of the PSF into segments}
 In the PSF, the direction-dependent scattering is described relative to the PSF center. AIA is suspected to scatter light in each direction over the full length of the detector. Therefore, we set the size of the PSF to twice the size of the detector to be able to describe this direction-dependent  long-range scattering originating from any point on the detector. Furthermore, we expect mostly isotropic optical aberrations and thus an isotropic scattering function, as AIA has an on-axis optical design. Deviations from this design, due to each channel utilizing only half of the respective mirror area, were found to be negligible (see below). In addition, we expect that the weights of the PSF coefficients decrease rapidly moving a few pixels away from the PSF center and then much slower towards the PSF tail. For these reasons,  we discretize the diffuse portion of the PSF into 50 shells:  10~shells from the PSF center to a distance of 5~px from the PSF center; then, 5 more shells to a distance of 10~px from the PSF center; and subsequently 35~shells with exponentially increasing shell widths up to the outer edge of the PSF. This discretization is shown in Figure~\ref{fig:tail_overview}(c).  Each shell  corresponds to one PSF coefficient to be fitted. We also experimented with anisotropic,  angular-resolved PSF discretizations to capture the slight deviations from the on-axis design. However, these deviations could not be resolved with the available data.

\paragraph{(d) Deconvolve the lunar transit images with the PSF}
To approximate the true intensities in the unocculted portion of the image and to diminish the scattered light in the lunar-occulted pixels, we deconvolve the images with the PSF. In the first iteration, we use only the known diffraction pattern PSF. In the subsequent iterations, we combine the known diffraction pattern PSF with the fitted diffuse scattered light PSF from the previous iteration. For the deconvolution, we use the basic iterative deconvolution (BID) algorithm and constrain the image reconstructions to positive intensities  \citep{cittert1931, iinuma1967a, iinuma1967b, jansson1968, jansson1970a, jansson1970b, hofmeister2023}. BID is able to retrieve photons that are scattered in the image plane out of the field of view of the detector and therefore gives robust image reconstruction results for PSFs that have a significant amount of long-distance scattered light \citep{hofmeister2023}.

\paragraph{(e) Identify the occulted pixels}
To identify the location of the occulted pixels using the PSF deconvolved image, we first fit a circle to the lunar edge using a Hough transformation \citep{hough1959}. This gives a rough approximation of the location of the lunar disk. To enhance the accuracy of the fitted lunar-solar boundary, we then reduce the radius of the fitted circle by 5~px and then extend the occulted region towards the solar disk until an intensity of \SI{20}{\percent} of the mean unocculted solar disk intensity is reached.  

\paragraph{(f) Approximate the true images}
To approximate the true image, we use the PSF reconstructed image and set the intensities within the lunar-occulted region to zero.

\paragraph{(g) Select pixels for the system of equations}
We have  $50$~PSF segments, i.e., $50$~PSF coefficients to fit, but typically more than one million occulted pixels per AIA image. As each pixel results in one line in the system of equations, the system is strongly overdetermined. Therefore, we can reduce the system of equations to a subset which contains the most reliable lines. (1) We discard all occulted pixels that are closer than 5 px to the lunar-solar boundary. As the lunar transit moves during each exposure, this removes occulted pixels that have potentially not been occulted for the full exposure time. (2) We discard all occulted pixels that are farther than 
\SI{1300}{\arcsec}, i.e., 2166~px, from the image center. This excludes all pixels close to the image corners, where vignetting is apparent. (3) We discard all occulted pixels that are farther than \SI{20}{\percent} of the image edge length from the lunar-solar boundary, i.e., 820~px. These pixels  are deep in the lunar occultation region, i.e., where the observed intensity is very low and thus where calibration artifacts are most prominent. (4) We discard all occulted pixels that are closer than $20$~px to the lunar-interplanetary space boundary. These pixels receive instrumental short- and medium-distance scattered photons from the low-intensity off-limb corona. Due to the low signal-to-noise ratio in the low-intensity off-limb corona, and thus the high uncertainty in the short-distance scattered light, these lines in the system of equations increase the uncertainty in the fit instead of constraining it. (5) We assign the remaining occulted pixels into $50$ image segments, depending on the distance to the lunar-solar boundary. The width of these segments are the same as the width of the shells in the discretized PSF, with the smaller segments close to the lunar-solar boundary and the larger segments far in the lunar-occulted region. The pixels in the smaller segments  close to the lunar-solar boundary receive mostly short-distance scattered light and thus mainly constrain the PSF core and wing, while the pixels in the segments far in the lunar occulted region only receive long-distance scattered light and thus constrain the PSF tail. These segments will be used for a stratified randomization, where drawing an equal number of occulted pixels from each segment will guarantee robust constraints for the fit of all PSF coefficients. We dub these segments the selection mask, which is shown in  Figure~\ref{fig:tail_overview}(g).
(6) We identify unocculted pixels within a distance of $5$ px of the lunar-solar boundary and add a small number of these unocculted pixels to the system of equation. This improves the fit between the simulated and observed scattered light across the lunar-solar boundary.

\paragraph{(h) Set up the system of equations}
As setting up the system of equations is computationally expensive, we further reduce the number of lines in the system of equations. For each AIA image, we draw randomly $600$~occulted pixels from each of the $50$~segment in the selection mask and \num{1000} unocculted pixels from the region at the lunar-solar boundary. This results in \num{31000} selected pixels per AIA image, i.e., \num{31000} lines in the system of equations per AIA image. To reduce the noise in the measured intensities of the selected occulted pixels, we replace the measured intensity of the occulted pixels by the nearby-neighbor-averaged intensity. We average over an area of $3\times 3$~pixels for occulted pixels with a distance of $15$~to $35$~px to the lunar-solar boundary, over $7\times 7$~pixels for distances of $35$~to $65$~px, over $13\times 13$~pixels for distances of $65$~to $105$~px, over $21\times 21$~pixels for distances of $105$~to $255$~px, and over $51\times 51$~pixels for distances of $>255$~px. For the occulted pixels with distances $<15$~px to the lunar-solar boundary and for the unocculted pixels, no averaging is performed. We derive for each selected pixel the coefficients $A_S$ and $B$ of Equation~\ref{eq:psf_px3}. Finally, we normalize each line of the system of equations by its observed intensity, as the observed intensities span several orders of magnitude from the illuminated edge to deep within the occulted region. With \numrange{8}{10} AIA images per AIA channel, this results in \numrange{248000}{310000} lines in the system of equations per AIA channel.

\paragraph{(i) Fit the PSF from the system of equations and (j) iterate }
In the system of equations defined by Equation~\ref{eq:psf_px3}, $I_\text{o,$\pmb{r}$}$, $A_S$, and $B$ are known. The PSF weights for the diffuse scattered light, $\widetilde{\text{psf}}_S$, are determined by a multi-linear fit. 

For the first three iterations of the algorithm, we do not parameterize the PSF   coefficients but directly fit them from the system of equations. For this, we randomly select \num{103000} lines of the system of equations:  \num{2500} lines from each of the $50$ selection mask segment and \num{3000} lines from the unocculted pixels along the lunar-solar boundary.
For the fit, we use a Monte Carlo walker method. The method iteratively converges to the solution by performing a random step in the parameter space and evaluating if the RMSE reduces. If the RMSE reduces, the step is accepted as new starting point in the parameter space; if not, it is rejected as a new starting point with a probability of \SI{80}{\percent} and accepted with a probability of \SI{20}{\percent}. By iteratively decreasing the step size, the algorithm converges toward the solution of the system of equations. Due to the stochastic nature, this method is more robust to overfitting than gradient-descend methods and does not require a regularization. 
We perform \num{20000} Monte Carlo steps to converge towards the fit solution, followed by \num{80000} steps to sample the immediate vicinity of the solution. The solution is expected to not be unique due to photon noise and the instrumental calibration uncertainties. Therefore, we average all accepted steps in the vicinity of the solution to obtain a robust fit result. We repeat this fitting procedure $10$~times, which reduces the effect of the choice of the selected lines, discard the $3$~solutions with the largest RMSE, and average the remaining $7$~solutions to obtain the final fitted PSF coefficients.  

In the last two iterations of the algorithm, we parametrize the PSF coefficients by 
\begin{equation}
    \widetilde{\text{psf}}_S(\Delta r) = a\ \frac{1}{b + \Delta r^c} + d\ \frac{1}{e + \Delta r^f}, \label{eq:psf_fit_tail_wing}
\end{equation}
which corresponds to a superposition of two Lorentzians.  In the case where~$b=0$ and $e=0$, the equation becomes equal the superposition of two power laws. The use of a superposition of two Lorentzians or two power laws, respectively, will be justified in the next section.

Having fitted the diffuse portion of the scattered light, we combine it with the PSF diffraction pattern to obtain the total PSF, as shown in Figure~\ref{fig:tail_overview}(i).


\subsection{Results}

First, we focused on the long-distance diffuse scattered light described by the PSF tail. Long-distance scattered light can be measured in regions far inside the lunar occultation region. The long-distance scattered light observed in these regions does not depend strongly on the quality of the lunar-solar boundary determination. Therefore, we neglect the movement of the moon during long exposures and use long-exposure images for their increased signal-to-noise ratio. We determine the most likely PSF tail coefficients and their uncertainty from these images by running the inversion algorithm $100$~times for each AIA channel, where we selected for each run $10$~long-exposure images with replacement from the pool of lunar-transit images. 

\begin{figure}
    \centering
    \includegraphics[width=\textwidth]{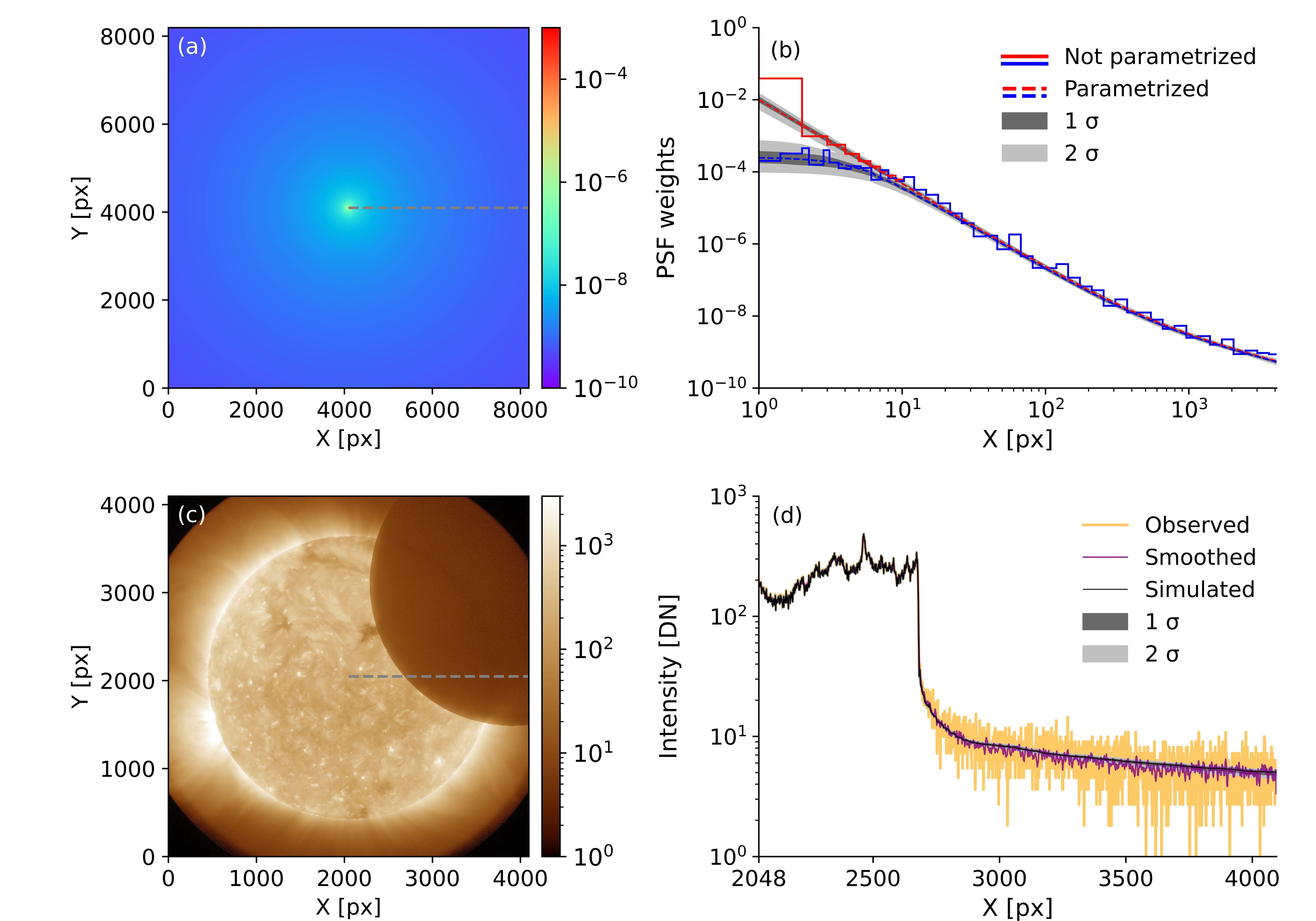}
    \caption{Representative evaluation of the AIA \SI{193}{\angstrom} fit using the 2010~October~7 lunar transit. (a) Fitted PSF from the long-exposure images. (b) Fitted PSF coefficients along the dashed gray line in (a). The blue solid line shows the unparameterized fit result from the long-exposure images and the blue dashed line the fit result using a parameterization of two Lorentzians.  The red solid line revises the fit using the short-exposure images and the red dashed line parameterizes the new fit by two power laws. (c) Observed lunar transit image. (d) Intensity profile along the dashed gray  line in (c). The orange line shows the observed intensities and the purple line the observed intensities smoothed by a $5 \times 5$ pixel kernel. The black line and the shaded areas show the median, $1\sigma$, and $2\sigma$ uncertainties.}
    \label{fig:tail_results1}
\end{figure}

A representative fit result for the \SI{193}{\angstrom} channel and the associated evaluation is shown in Figure~\ref{fig:tail_results1}. Panel~(a) shows the fitted PSF for the long-distance scattered light, and the blue lines in panel~(b) show these fitted PSF weights along a horizontal slice through the PSF center. The solid line, representing the unparameterized fit, reveals a long PSF tail with light scattered over the full image width. The dashed line, representing the parameterized fit using 2~Lorentzians, follows well the non-parametrized fit. The fitted PSF tail coefficients shown here are robust as they primarily depend on the intensities far into the lunar-occulted regions. However, the apparent flattening seen in the PSF wing coefficients should not be considered robust, as this region of the PSF depends on the intensities close to the lunar-solar boundary where the moon has moved during the exposure.

To remedy this effect, we revise the fit using the short-exposure lunar-transit images. In these images, the moon moved by \SI{<1}{px} during the exposure. We repeat the fit for the PSF wing coefficients using these images, while we keep fixed the PSF tail coefficients from the previous fit.  The revised fit result is shown by the red lines Figure~\ref{fig:tail_results1} (b). The revised fit without parameterization, i.e., the red dashed line, shows that  PSF weights do not flatten towards the PSF center. Instead, they now steadily increase towards the PSF center, apparently following a power law. Therefore, we re-parameterized the fitted PSF weights accordingly by setting the coefficient~$b$ and~$e$ in Equation~\ref{eq:psf_fit_tail_wing}  to~zero. The result is shown by the red dashed line in Figure~\ref{fig:tail_results1} (b). 

Next, we evaluated the quality of our fit. Figure~\ref{fig:tail_results1}~(c) shows a representative evaluation for the accuracy of the fitted diffuse scattered light, using the 2014~September~24 lunar transit image. To evaluate the accuracy, we simulated the light scattered into the lunar-occulted region  and compared it to that observed. We began with approximating the true image by deconvolving the observed image with the PSF containing the diffuse scatted light and the diffraction pattern and then setting all intensities within the occulted image region to zero.
Next, we simulated the instrumental scattered light by convolving the approximated true image with this PSF. 
The amount of simulated scattered light into the occulted region is a robust quantity and only slightly dependent on the quality of the reconstructed illuminated solar disk. It is mainly sensitive to the PSF used, but largely independent on the quality of the reconstructed true solar disk intensities on small scales. Figure~\ref{fig:tail_results1}~(d) shows a comparison between the simulated intensities in the lunar-occulted regions and those observed. Since large differences between the simulated and observed intensities are not apparent, we conclude that the fitted PSF weights are accurate.



Finally, in Figure~\ref{fig:tail_results2} and~\ref{fig:tail_results2_cont}, we present the fit results for all AIA channels. Figure~\ref{fig:tail_results2}~(a) shows the amount of diffuse scattered light per AIA channel. We find that AIA diffusely scatters $10$~to~\SI{35}{\percent} of the collected light over short, medium, and large distances, which has not yet been accounted for by the diffraction pattern.  For each AIA telescope, we find that the short-wavelength channel scatters more light than the long-wavelength channel. Furthermore, the uncertainty in the fitted amount of diffuse scattered light is on average much lower in the channels that observe stronger quiet Sun emission lines ($171$, $193$, $211$, and \SI{304}{\angstrom}) as compared to the channels that observe weaker quiet Sun emission lines ($94$,~$131$ and~\SI{335}{\angstrom}). This emphasizes the importance of a good signal with minimal calibration artifacts for PSF fitting.
The remaining panels show the PSF weights describing diffuse scattered light for each AIA channel on the left side, and the associated percentage of light that is scattered farther than a given distance on the right side. The PSF parameterizations are given in Table~\ref{tab:fitted_tail_parameters}.

\begin{figure}
    \centering
    \includegraphics[width=\textwidth]{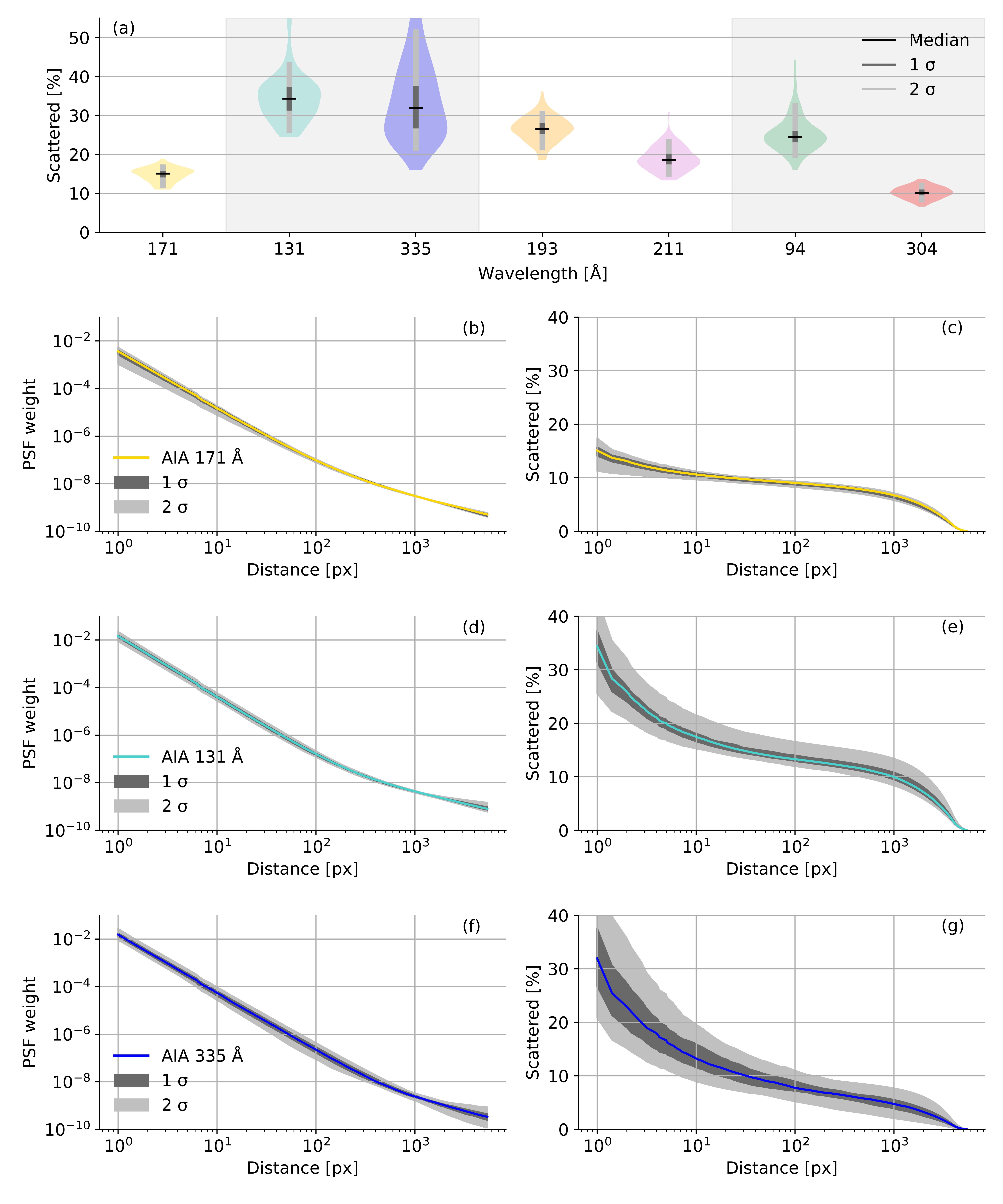}
    \caption{Results for the diffuse scattered light for all AIA channels, derived from $100$~runs per AIA channel. (a) Amount of scattered light.  Each of the vertical gray and white areas mark channels on the same AIA telescope. For each AIA channel, the color-shaded areas show the probability distribution for the scattered light as derived from all lunar occultations. The horizontal bar corresponds to the median value. The dark and light gray boxes are the $1\sigma$ and $2\sigma$ ranges, respectively, of the probability distributions.  (b),~(d), and (f): Fitted PSF weights for the~$171$,~$131$, and \SI{335}{\angstrom} channels. (c),~(e), and~(g): The corresponding amount of light that is scattered farther than a given distance vs. the distance.}
    \label{fig:tail_results2}
\end{figure}

\begin{figure}
    \centering
    \includegraphics[width=\textwidth]{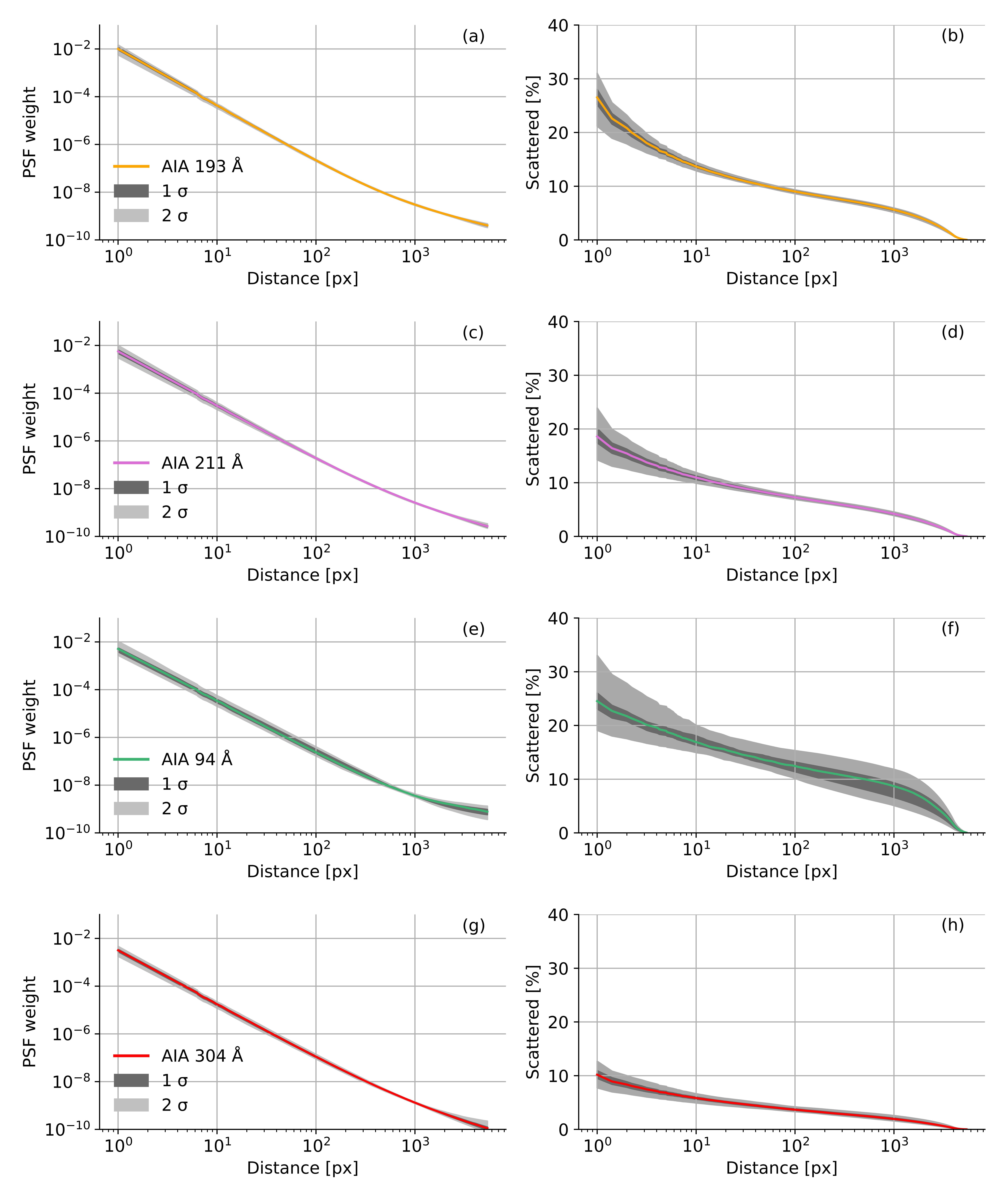}
    \caption{Continuation of Figure~\ref{fig:tail_results2} but for the AIA~$193$,~$211$,~$94$, and \SI{304}{\angstrom} channels.}
    \label{fig:tail_results2_cont}
\end{figure}

\begin{table}[t]
    \centering
    \caption{Final PSF parameters of Equation~\ref{eq:psf_fit_tail_wing} for the diffuse scattered light.}
    \label{tab:fitted_tail_parameters}
    \begin{tabularx}{.73\textwidth}{c | r r r r r r c } 
        AIA channel & \multicolumn{1}{c}{a} & \multicolumn{1}{c}{b} & \multicolumn{1}{c}{c} & \multicolumn{1}{c}{d} & \multicolumn{1}{c}{e} & \multicolumn{1}{c}{e} & \multicolumn{1}{c}{Diffuse scattered light} \\
        \midrule
        \SI{171}{\angstrom} & 3.65e-3(127) & 0 & 2.33(9) & 2.09e-6(329) & 0 & 0.96(12) & 15.5(18)\,\% \\
        \SI{131}{\angstrom} & 1.47e-2(50) & 0 & 2.49(10) & 2.56e-6(360) & 0 & 0.94(21) & 34.4(60)\,\% \\
        \SI{335}{\angstrom} & 1.70e-2(58) & 0 & 2.47(13) & 5.06e-6(410) & 0 & 1.13(36) & 32.5(96)\,\% \\
        \SI{193}{\angstrom} & 1.05e-2(28) & 0 & 2.35(7) & 2.85e-6(276) & 0 & 1.03(12) & 26.9(33)\,\% \\
        \SI{211}{\angstrom} & 5.90e-3(225) & 0 & 2.27(8) & 8.60e-6(341) & 0 & 1.22(13) & 18.9(31)\,\% \\
        \SI{94}{\angstrom}  & 5.62e-3(289) & 0 & 2.32(7) & 5.06e-6(145) & 0 & 1.04(18) & 23.1(48)\,\% \\
        \SI{304}{\angstrom} & 3.16e-3(93) & 0 & 2.22(6) & 1.93e-6(391) & 0 & 1.15(29) & 10.3(15)\,\% \\
    \end{tabularx}
\end{table}

\section{The PSF core}

\begin{figure}
    \centering
    \includegraphics[width=1.\linewidth]{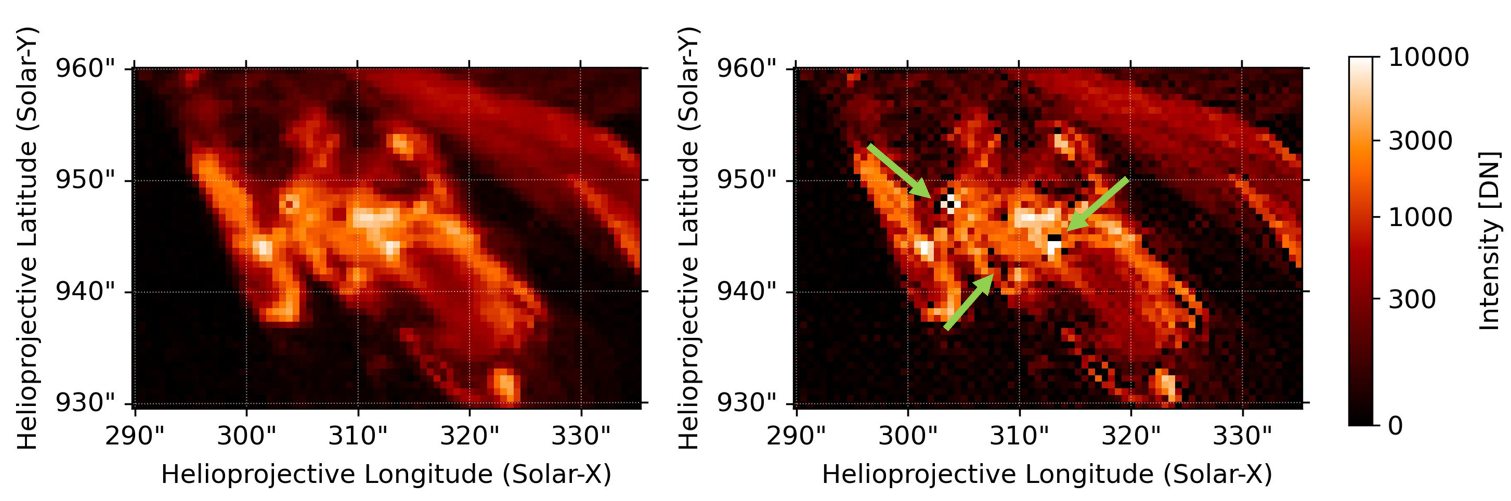}
    \caption{Flare observed by AIA~\SI{304}{\angstrom} on 2023~March~3. Left: Original image. Right: Deconvolved image assuming a PSF core with a width of~$\sigma=\SI{0.63}{px}$. Green arrows point to reconstruction artifacts, i.e., pixels with negative values, due to too large of a chosen PSF core width.}
    \label{fig:flare_max_example}
\end{figure}

\begin{table}[]
    \centering
    \begin{tabular}{l| r r r r r r r}
         & \multicolumn{7}{c}{AIA channel} \\
         &  \SI{94}{\angstrom} & \SI{131}{\angstrom} & \SI{171}{\angstrom} & \SI{193}{\angstrom} & \SI{211}{\angstrom} & \SI{304}{\angstrom} & \SI{335}{\angstrom} \\ \toprule
         $\sigma_\text{max}$ [px]& $0.59$& $0.55$& $0.44$& $0.46$& $0.51$& $0.48$& $0.45$    \end{tabular}
    \caption{Upper limit on the PSF core width, assuming a Gaussian core.}
    \label{table:coresizes}
\end{table}

The PSF core affects the sharpness of small-scale solar features. The width of the PSF core depends on the size of the Airy disk arising from diffraction from the primary mirror, manufacturing uncertainties of the optical components, the alignment accuracy of the optical components, CCD charge spreading, jitter residuals from the pointing of the instrument, and on-orbit thermal effects. 

The width of the PSF core is best determined from observations of a point source, such as a distant star: the blurring in these observations can be attributed to the PSF core. Unfortunately, for AIA, such observations are not available. Without these, the determination of the PSF core width is very challenging. We attempted to derive a reliable estimate of the PSF core width, but were only able to  determine an upper limit. In the following, we summarize our efforts, in the expectation that this will help to design a more reliable calibration for future instruments.

Without a point source, the calibration of the PSF core can be based on extremely bright, extended sources such as solar flares, on partially occulted regions due to lunar or planet transits, or on blind-deconvolution algorithms. In the following, we model the PSF core as a Gaussian with a~$\sigma$ that we fit for.  We refer to the total PSF as the PSF core convolved with the combined PSF of the diffraction pattern  (Section~\ref{sec:diffraction}) and the diffuse scattered light (Section~\ref{sec:diffuse}). 

\subsection{Solar flares}
Solar flares are spatially small but extremely bright. Therefore, the PSF core dominates the scattered light close to the flare; the diffraction pattern is clearly visible at larger distances; and the diffuse scattered light only creates a small background that can be neglected here. 
The first approach using solar flares focuses on determining an upper limit for the width of the PSF core. This can be achieved by deconvolving flare images with various PSFs of increasing core width~$\sigma$  and analyzing at which value of~$\sigma$ that artifacts begin to appear. We define artifacts as  clusters of image pixels with negative intensities. For this approach, we use BID as the deconvolution algorithm and allow negative pixel intensities in the image reconstruction process. An example is shown in Figure~\ref{fig:flare_max_example}. Panel~(a) shows the original image and panel~(b) the deconvolved image with apparent artifacts due to too large of a PSF core width. By systematically varying~$\sigma$ and visually inspecting all flare images in our study, we found an upper limit for the core widths of  $\sigma = 0.44-0.59$~px. The upper limit for each channel is given in Table~\ref{table:coresizes}. These core widths imply that  $<6-11$\si{\percent} of light, depending on the AIA~channel, is scattered to each horizontally and vertically adjacent pixel, that $<0.5-2.6$\si{\percent} of light is scattered to each diagonally adjacent pixel, and that the light scattered over larger distances by the PSF core is negligible. This constrains the total amount of scattered light by the PSF core to the immediately adjacent pixels to  $<25-54$\si{\percent}.

A second approach involving flare images is to analyze the diffraction pattern of the flares. The PSF core is imprinted in each diffraction peak, and so, the spatially resolved diffraction peaks should allow for an estimate of the core width. If the object being imaged were a point source, the image pixels adjacent to the diffraction peaks would directly measure the PSF core. But AIA spatially resolves the flares, so a more sophisticated approach has to be taken. The methodology that we used follows the calibration procedure for the diffraction pattern from Section~\ref{sec:diffpatterncal}; but instead of fitting for the mesh parameters, we fit for the PSF core width. 
First, we remove the quiet Sun background from the flare image by subtracting a prior non-flare image. Then, we approximate the true image of the flare by deconvolving this image with the fitted PSF from Sections~\ref{sec:diffraction} and~\ref{sec:diffuse} convolved with a Gaussian test PSF core. We remove the observed diffraction pattern by setting all intensities outside the flare to zero. Finally, we simulate the scattered and diffracted light by convolving this image with the PSF.  By varying the PSF core width and comparing the simulated diffraction pattern with that observed, we should be able to determine the PSF core width. 

To test the robustness of this approach, we  created a synthetic true flare image consisting of a small cluster of bright pixels and assumed the PSF to have a core width of $\sigma=\SI{0.5}{px}$. We then derived a synthetic observed image by convolving this image with a PSF consisting of the Gaussian core, the diffraction pattern, and the diffuse scattered light. Then, we attempted to re-determine the width of the PSF core using the approach described above, while also considering the uncertainties in the mesh parameters (Table~\ref{table:coresizes}). When neglecting the uncertainties in the mesh parameters, we were able to accurately derive the PSF core width. However, when assuming an error in the angle of the diffraction pattern of  only \SI{0.05}{\degree}, the fitted PSF core width became random. As the accuracy of our calibration of the diffraction pattern angle is this order of magnitude, any fitted PSF core width determined using this approach is unreliable. 

A third approach is to vary the PSF core width, to deconvolve flare images with the assumed PSF, and to visually check which PSF core width best removes the diffraction pattern from the images. The drawback of this approach is again that the removal of the diffraction pattern depends on an accurate determination of the diffraction pattern PSF. Since the precise location of the diffraction pattern peaks varies slightly between  flare images, due to mechanical and temperature deformations of the entrance filter meshes, we have to assume that our static diffraction pattern PSF fitted in Section~\ref{sec:diffraction} will show small errors in the location of the diffraction pattern peaks, errors that are dependent on the specific flare image under consideration.  Therefore, the PSF core width for the image where the diffraction pattern best vanishes has corrected for both the true PSF core and for errors in the diffraction pattern location. This makes this approach unreliable to determine the PSF core width.

\subsection{Partial occultations} \label{subsec:core_partial}
In partially occulted images, such as from lunar or planet transits, it is known that the true intensities in the occulted pixels are zero. Thus, any residual intensities observed in the occulted pixels are related to instrumental scattering by the diffraction pattern, the diffuse scattered light, and the PSF core.  This offers another approach to determine the PSF core.

Initially, we tried to determine the width of the PSF core using the methodology of Section~\ref{sec:diffuse}, i.e., we assume that the intensity in the occulted pixels adjacent to the occultation edge arises from  the PSF core. From the known contribution of the diffraction pattern and the diffuse scattered light, one can potentially derive the remaining intensity due to scattering by the PSF core and the PSF core width. However, this approach had three drawbacks. First, the contributions from the fitted diffracted and diffuse scattered light and their uncertainties (Table~\ref{tab:fitted_mesh_parameters} and~\ref{tab:fitted_tail_parameters}) have to be considered.  The amount of light that is diffracted and diffusely scattered to a neighboring pixel has an absolute uncertainty of up to \SI{10}{\percent} of the collected light, primarily due to the uncertainty in the diffuse scattered light. This is comparable to the amount of light that is scattered by the PSF core to an adjacent pixel, which is $<6-11$\si{\percent} of the collected light. Therefore, the uncertainty in the fitted diffuse scattered light contribution will have a large effect on the to-be-derived PSF core.
Second, one needs to reliably determine the occultation edge with an accuracy \SI{<1}{px}, i.e., we have to know exactly which pixels are fully occulted, partially occculted, and fully illuminated. Since the occulted region moves during each exposure, it is difficult to reliably determine this.  Third, as noted in Section~\ref{sec:instrumental_issues}, the calibration in the images shows systematic errors at the order of \SI{0.2}{DNs} for the read-out amplifiers. For the low-intensity channels,  such as the AIA~$94$, $131$, and~\SI{335}{\angstrom} channels, this systematic error is significant relative to the expected amount of light that is scattered by the PSF core to an adjacent pixel (which is \SI{<1}{DNs}). Due to the above issues,  and dependent on the lunar transit image used, we found PSF core widths of either  $\sigma=\SI{0}{px}$ or $\sigma \gg \SI{1}{px} $. A core width of~\SI{\gg1}{px} is larger than the upper limit on the core width  determined from the flare images. Therefore, we deem this approach unreliable.

We next tried to circumvent the uncertainty in the location of the occultation edge by analyzing the intensity profile across the occultation edge (instead of analyzing pre-selected pixels). To determine~$\sigma$, we varied the PSF core width, deconvolved the entire image with the full PSF, and checked at which~$\sigma$ the intensity close to the occultation edge became negative. For this approach, we allowed negative pixel intensities in the image reconstruction process by BID. However, we obtained ringing artifacts on both sides of the occultation edge in the image reconstructions. The ringing artifacts arise due to the steep intensity gradient at the occultation edge in combination with having had to allow for negative intensities and prevents us from robustly determining the width of the PSF core.

\subsection{Blind deconvolutions}
Blind deconvolution techniques aim to reconstruct the PSF simultaneously with the true image \citep{kundur1996}. Blind deconvolution techniques typically require priors to become stable. The most commonly used prior is to assume ``natural'' looking images in which objects have sharp edges. This assumption is mathematically described by a sparsity of the image gradient matrix. However, as AIA observes line-of-sight integrated emission, we expect fuzzy edges in solar images. Therefore,  such a prior cannot be used for solar images. 

Another prior is the information that the intensity in the occulted pixels should be zero. This was used by 
\citet{gonzalez2016}, who tried to determine the AIA~\SI{193}{\angstrom} PSF from Venus transit images. This approach has similar issues as mentioned in Section~\ref{subsec:core_partial}. Since planetary occultations in the field of view of AIA are small and very rare, there is not sufficient data to statistically resolve issues regarding the uncertainty in the AIA calibration and the location of the occultation edge. 

Finally, we tried the original approach of \citet{fish1995}, performing a blind deconvolution of AIA images based on the Richardson-Lucy algorithm with the prior that the PSF core is Gaussian. This gave a delta function for the PSF core. This indicates that the blind deconvolution algorithm did not converge to the correct solution. 

\subsection{Future outlook}
With the current methods and data, neither solar flares, lunar occultations, nor blind deconvolution can provide a precise value for the PSF core width. But we believe that it might be possible to resolve this issue in the future using either very long exposure images of lunar occultations, which diminish the error in the instrumental calibration and in the fitted diffuse scattered light, or sophisticated deep learning techniques, which fit the diffraction pattern simultaneously with the PSF core and allow for uncertainties in the location of the individual diffraction peaks.

\section{The revised PSFs of AIA}
In the following sections, we first present the derived PSFs and describe how to apply them to deconvolve AIA images. Then, we evaluate the quality of the fitted PSFs and present their effects on AIA images and associated DEM reconstructions.

\subsection{The PSFs}

\begin{figure}
    \centering
    \includegraphics[width=1.\linewidth]{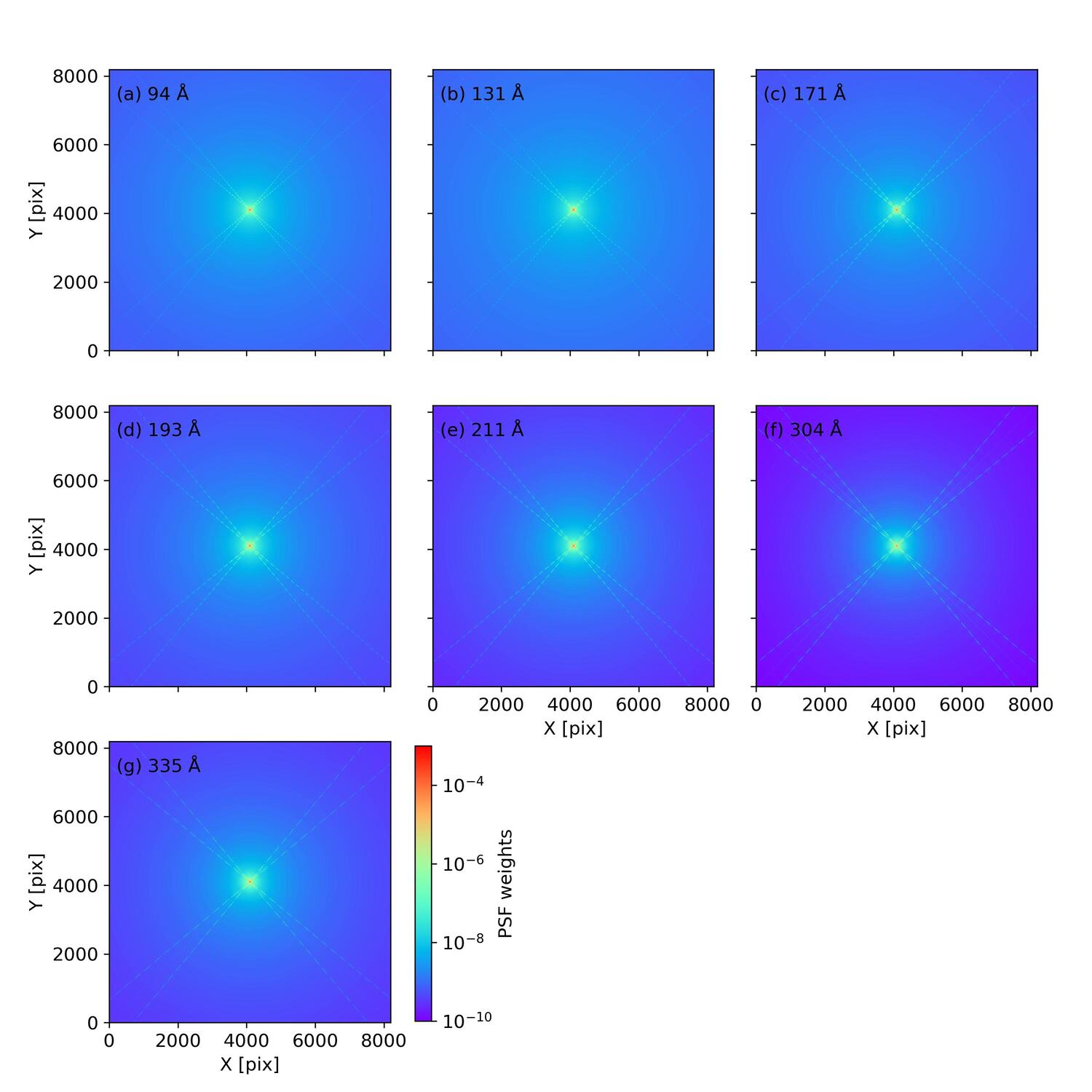}
    \caption{Revised PSFs for the 7 EUV AIA channels: (a) AIA \SI{94}{\angstrom}, (b) AIA \SI{131}{\angstrom}, (c) AIA \SI{171}{\angstrom}, (d) AIA \SI{193}{\angstrom}, (e) AIA \SI{211}{\angstrom}, (f) AIA \SI{304}{\angstrom}, and (g) AIA \SI{335}{\angstrom}. }
    \label{fig:allpsfs}
\end{figure}

\begin{figure}
    \centering
    \includegraphics[width=1.\linewidth]{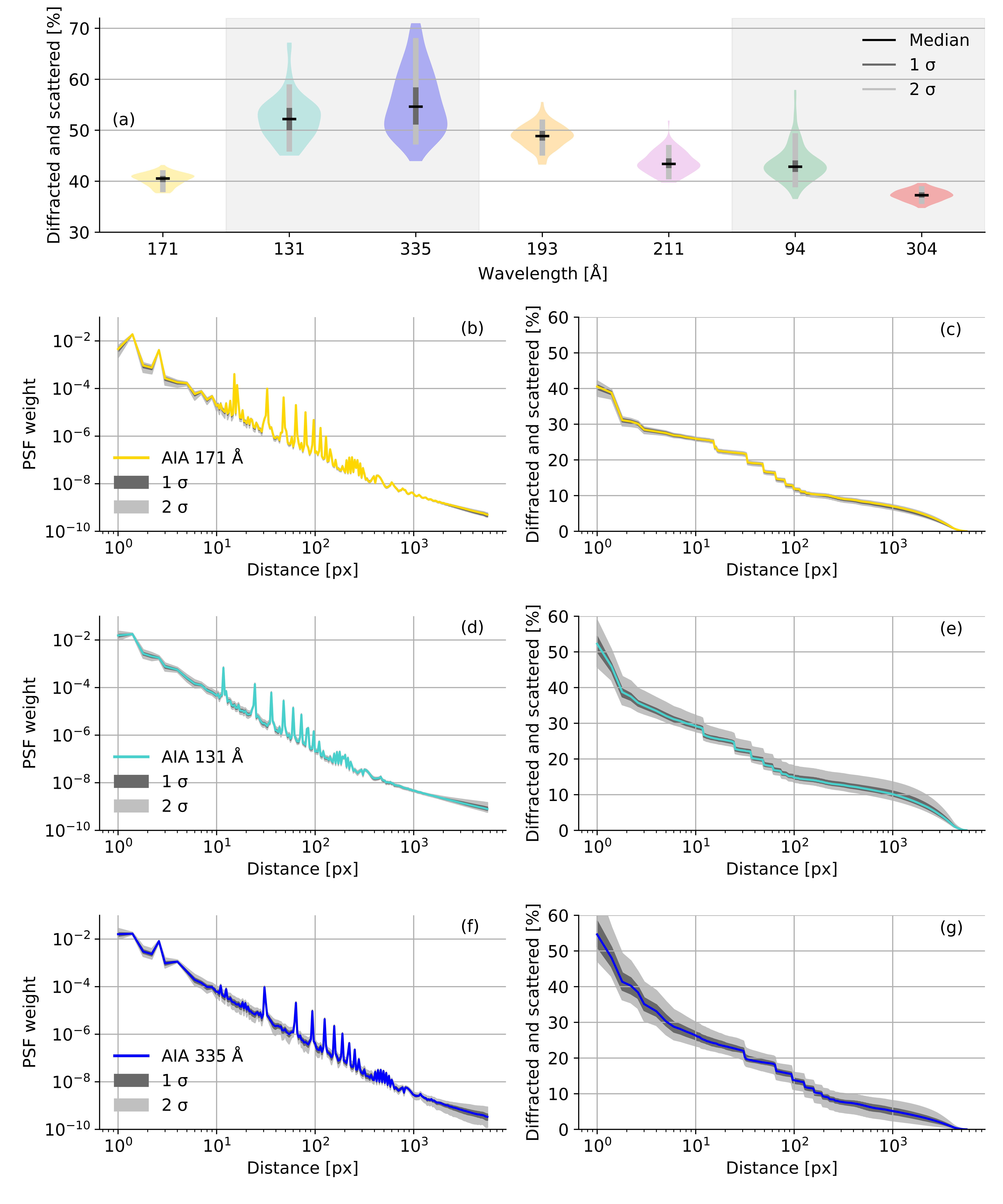}
    \caption{Our new PSFs for all AIA channels. (a) Amount of diffracted and scattered light. The color-shaded areas show the probability distribution for the diffracted and scattered light. Each of the vertical stripes mark channels on the same AIA telescope. Left panels: Fitted PSF weights for the~$171$,~$131$, and~\SI{335}{\angstrom} channels. Right panels: Amount of light that is diffracted and scattered farther than a given distance vs. the distance.}
    \label{fig:allpsfs_scatter1}
\end{figure}
\begin{figure}
    \centering
    \includegraphics[width=1.\linewidth]{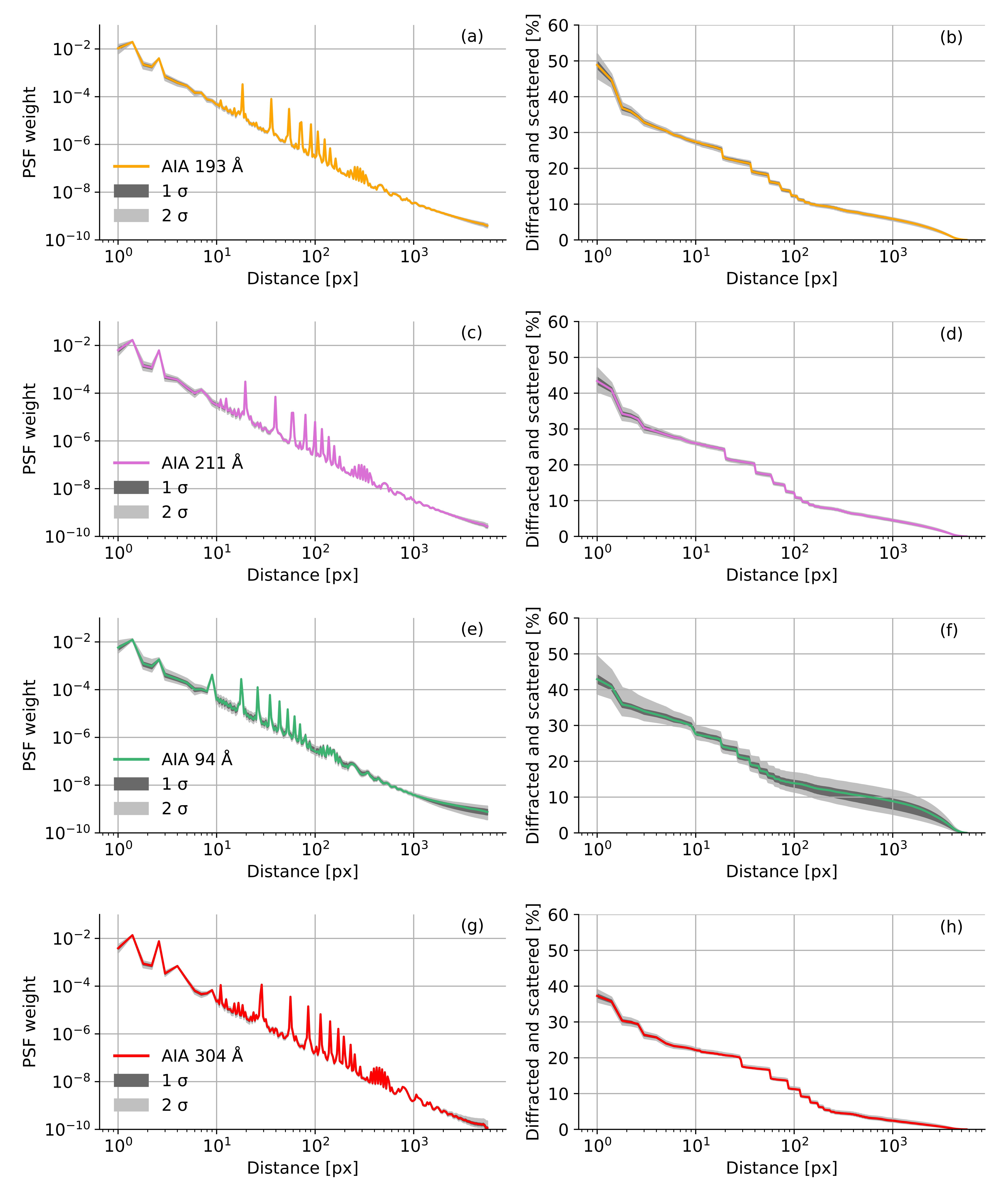}
    \caption{Continuation of Figure~\ref{fig:allpsfs_scatter1} for the AIA~$193$,~$211$, $94$,~and~\SI{304}{\angstrom} channels.}
    \label{fig:allpsfs_scatter2}
\end{figure}

We assemble our new AIA PSFs by merging the PSF of the diffraction patterns  from Section~\ref{sec:diffraction} with the PSF of the diffuse scattered light from Section~\ref{sec:diffuse}:
\begin{equation}
\text{psf}_{\text{new}} =\left( 1 - \text{psf}_{\text{scat},\Delta \pmb{r} = 0} \right)\ \text{psf}_{\text{diff}}   + \text{psf}_{\text{scat}}, 
\end{equation}
where $\text{psf}_{\text{new}}$ is the revised PSF, $\text{psf}_{\text{diff}}$ is the PSF of the diffraction pattern, and $\text{psf}_{\text{scat}}$ is the PSF of the diffuse scattered light. The summation here mimics a rough approximation for the convolution operation for instrumental PSFs,  for which the PSF center coefficient is much larger than all other PSF coefficients. We use this approximation here for consistency in our methodology, as the same approximation was implicitly used while fitting the PSF of the diffuse scattered light. We decided to not add an additional PSF core, as we believe that it is currently not possible to determine its width precisely. Therefore, the revised PSFs will slightly underestimate the sharpness in the reconstructed images for small-scale features with diameters smaller than a few pixels.

Figure~\ref{fig:allpsfs} shows the final revised PSFs for the 7 AIA EUV channels. They contain the diffraction pattern from the entrance and focal plane mesh filters and the diffuse scattered light. With a dimension of $8192 \times 8192$~pixels, i.e., double the size of the AIA images, they describe diffracted and scattered light over the full length of the detector. 

In Figures~\ref{fig:allpsfs_scatter1} and~\ref{fig:allpsfs_scatter2}, we present the properties of these new PSFs. Figure~\ref{fig:allpsfs_scatter1} (a) shows that about $40$~to \SI{60}{\percent} of light is diffracted and scattered in the AIA telescopes over the entire detector, resulting in a blurring and a reduction of the dynamic contrast of AIA images. The \SI{304}{\angstrom} channel diffracts and scatters a total of \SI{37}{\percent}, the smallest amount of light for any of the AIA channels. This is followed by the \SI{171}{\angstrom} channel with \SI{41}{\percent}, the \SI{94}{\angstrom} and \SI{211}{\angstrom} channels with \SI{43}{\percent}, the \SI{193}{\angstrom} channel with \SI{49}{\percent}, the \SI{131}{\angstrom} channel with \SI{52}{\percent}, and the \SI{335}{\angstrom} channel with \SI{55}{\percent}.
The remaining panels on the left side show the cylindrically averaged PSF weights as a function of distance to the PSF center. The panels on the right side show the amount of light that is diffracted and scattered farther than a given distance. Depending on the AIA channel, $23$~to~\SI{29}{\percent} of light is scattered farther than \SI{10}{px},  $11$~to~\SI{15}{\percent} farther than \SI{100}{px}, and $3$~to~\SI{10}{\percent} farther than \SI{1000}{px}.

\subsection{How to apply PSF deconvolutions}

Our revised PSFs are publicly available at \citet{Hofmeister2024_psfs}\footnote{The AIA PSFs can be downloaded from \url{https://doi.org/10.7910/DVN/DYT4ZL}}. They do not contain an additional PSF core, as we believe that the width of this core currently cannot be reliably determined. If desired, the user can add a custom PSF core to our PSFs. To do so, one would need to convolve our PSFs with an array that contains the desired PSF core.

Due to the significant amount of long-distance scattered light in the AIA channels, a portion of the light is scattered out of the field of view of the detectors. During the image reconstruction process, this light has to be retrieved to achieve correct results. As the Richardson-Lucy deconvolution algorithm \citep{richardson1972, lucy1974} conserves flux within the image, it cannot retrieve these photons and thus should not be used to deconvolve AIA images with our PSFs. A better alternative is the BID algorithm described in \citet{hofmeister2023}, which has a slightly worse noise performance but can retrieve these photons and thus avoids systematic errors.

The BID algorithm performs one step in the deconvolution procedure in the Fourier domain. All deconvolution operations in the Fourier domain imply periodic boundary conditions, i.e., they treat light which is scattered out at one image edge as incoming scattered light at the opposite image edge.  To break these periodic boundary conditions, one has to pad the region surrounding the images with zeros, where the extent of the padding in each direction has to be at least quarter of the length of the PSF edge. Therefore, for AIA image deconvolution, we have to deconvolve a padded $8192\times8192$~pixels AIA image with this new AIA PSF. To execute such a deconvolution operation within a reasonable time, we recommend to perform the deconvolution using a GPU, to perform a subimage deconvolution of the subfield of interest, or to rebin the images and the PSFs to a lower resolution. The associated required tools are described in \cite{hofmeister2023}\footnote{The software tools are available at \url{https://github.com/stefanhofmeister/PSF-Tools}}.

\subsection{Evaluation of the fitted PSFs}

\begin{figure}
    \centering
    \includegraphics[width=1.\linewidth]{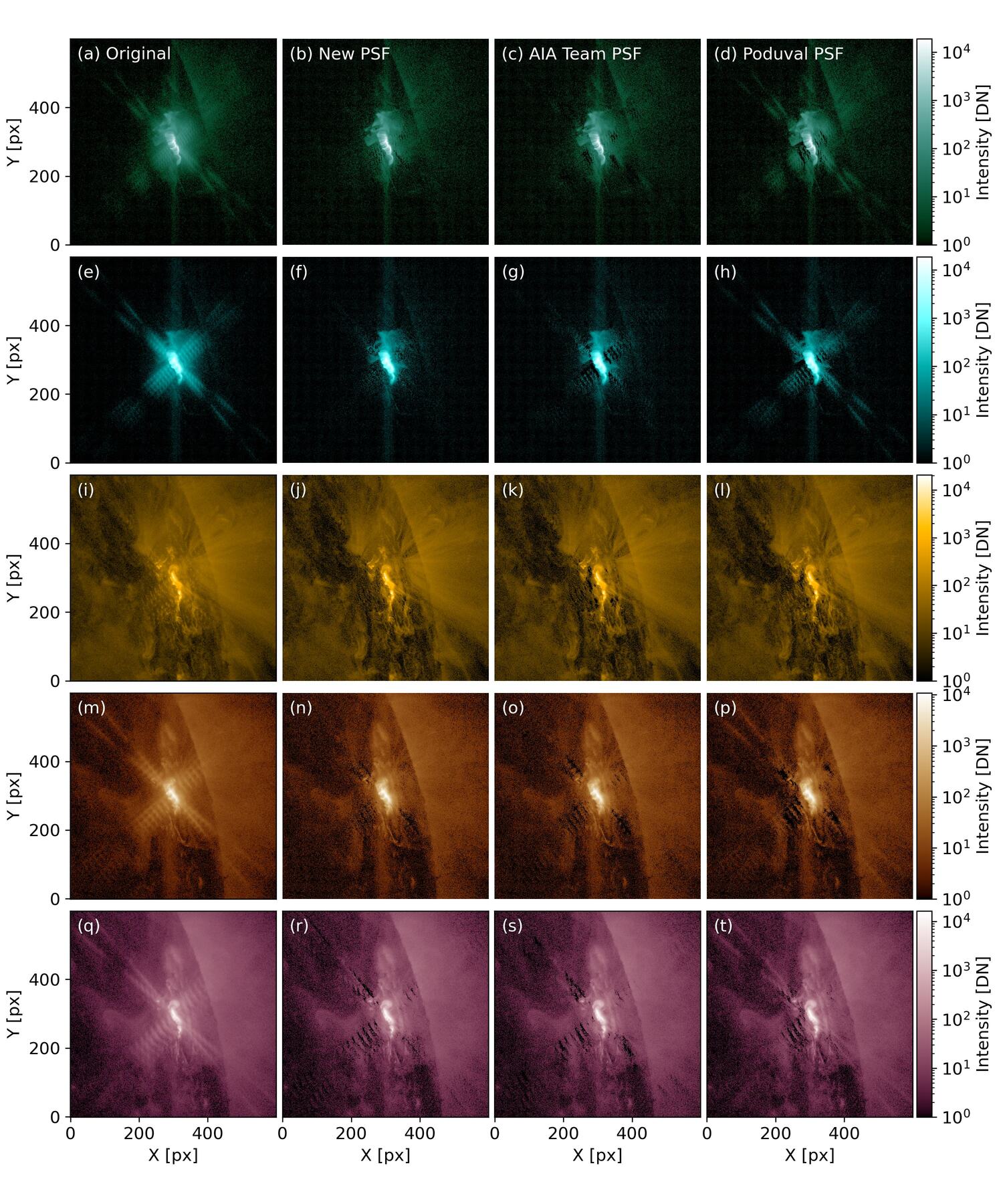}
    \caption{Evaluation of the new PSFs using 2011~September~8 images of an X~9.96 class flare. From left to right: Original images, deconvolved images using our new PSFs, the AIA team PSFs, and the Poduval PSFs. From top to bottom: Deconvolution results for the AIA~$94$,~$131$, $171$,~$193$, and~\SI{211}{\angstrom} channels.}
    \label{fig:eval_flares1}
\end{figure}
\begin{figure}
    \centering
    \includegraphics[width=1.\linewidth]{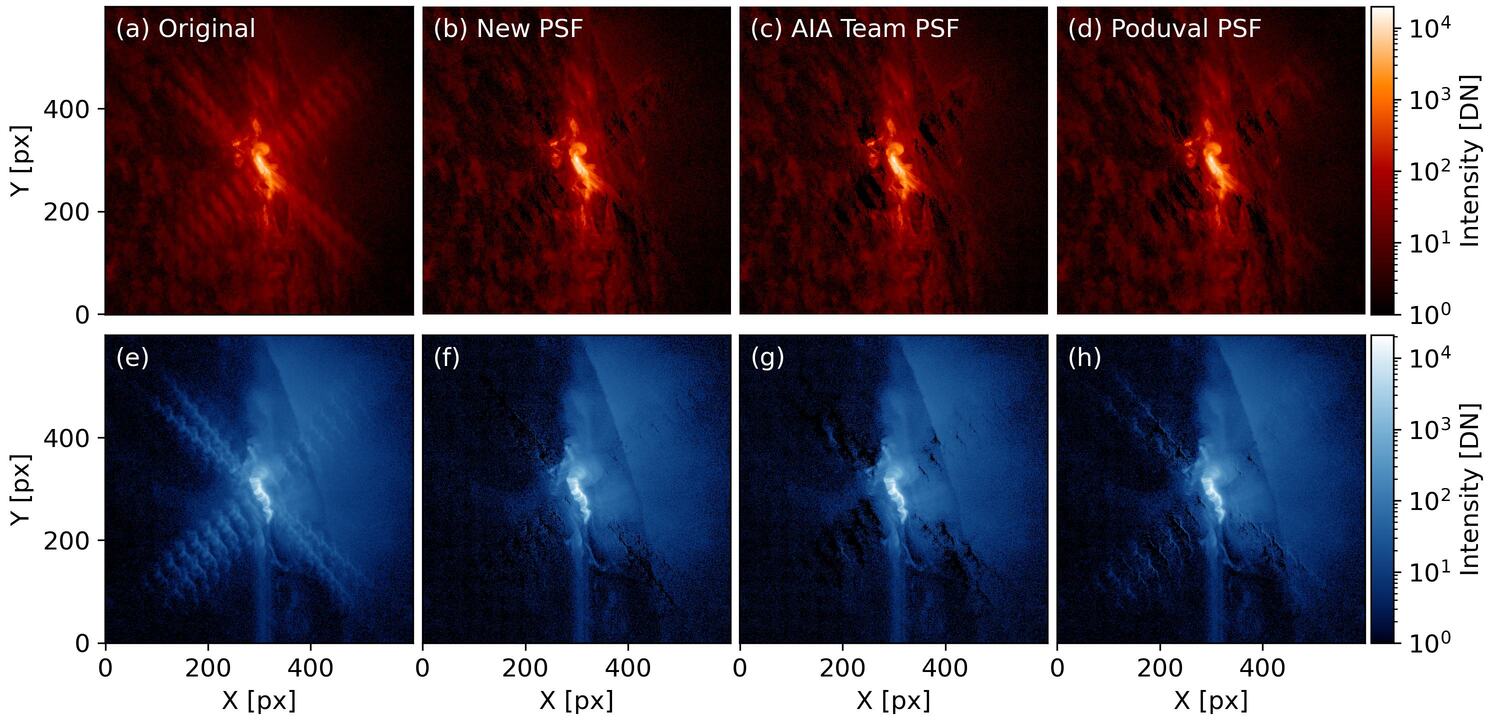}
    \caption{Continuation of Figure~\ref{fig:eval_flares1} for the AIA~$304$ and~\SI{335}{\angstrom} channels.}
    \label{fig:eval_flares2}
\end{figure}

\begin{figure}
    \centering
    \includegraphics[width=1.\linewidth]{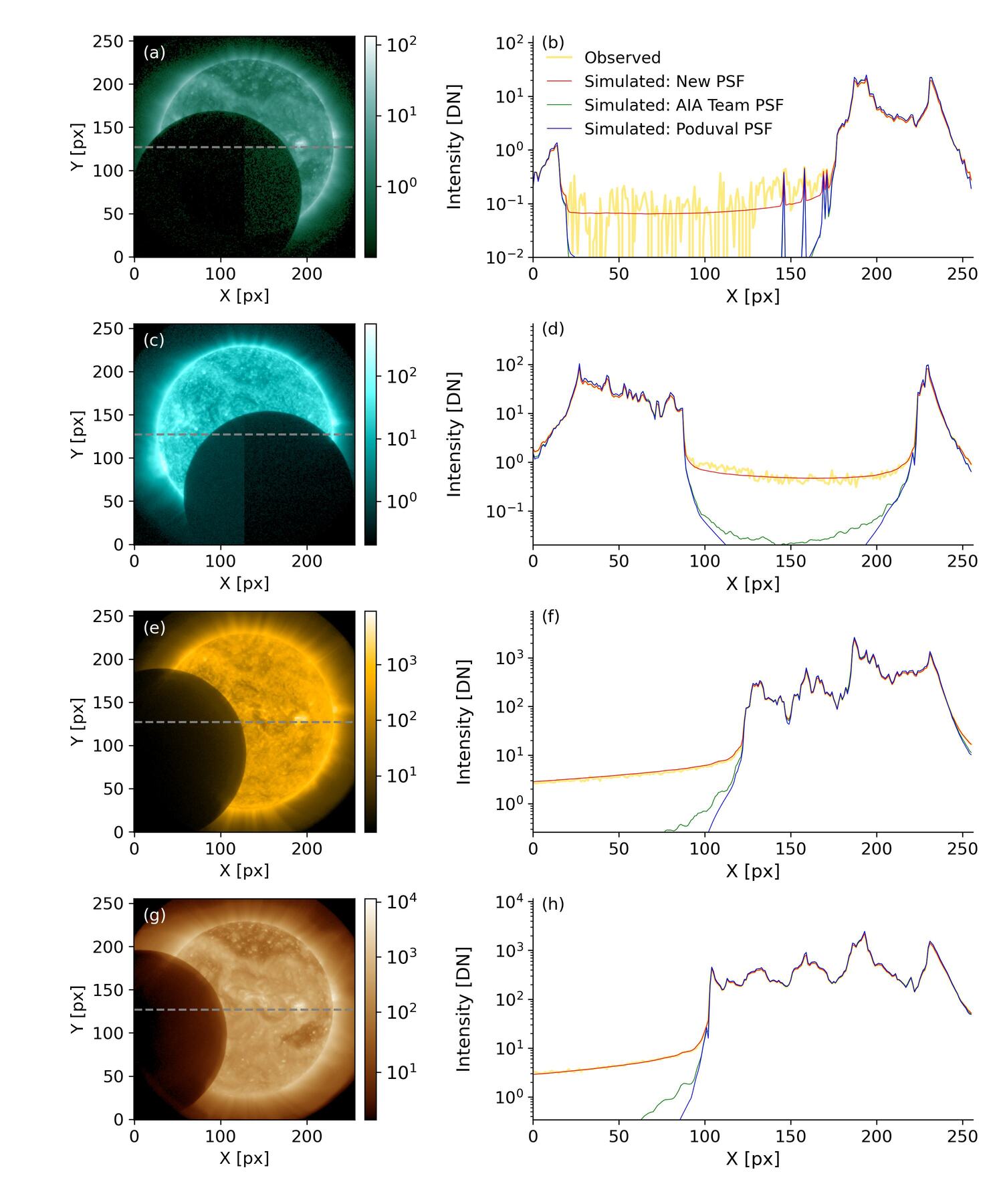}
    \caption{Evaluation of the new PSFs using the 2016~October~30 lunar eclipse images. Left: Eclipse image. Right: Intensity profile along the gray dashed line in the left image. The yellow lines are the intensities in the original AIA images, the red lines show the deconvolved intensities using our revised PSF, the green lines use the AIA team PSF, and the blue lines use the Poduval PSF. The rows from top to bottom show the results for the AIA~$94$,~$131$, $171$,~and \SI{193}{\angstrom} channels.}
    \label{fig:eval_eclipse1}
\end{figure}
\begin{figure}
    \centering
    \includegraphics[width=1.\linewidth]{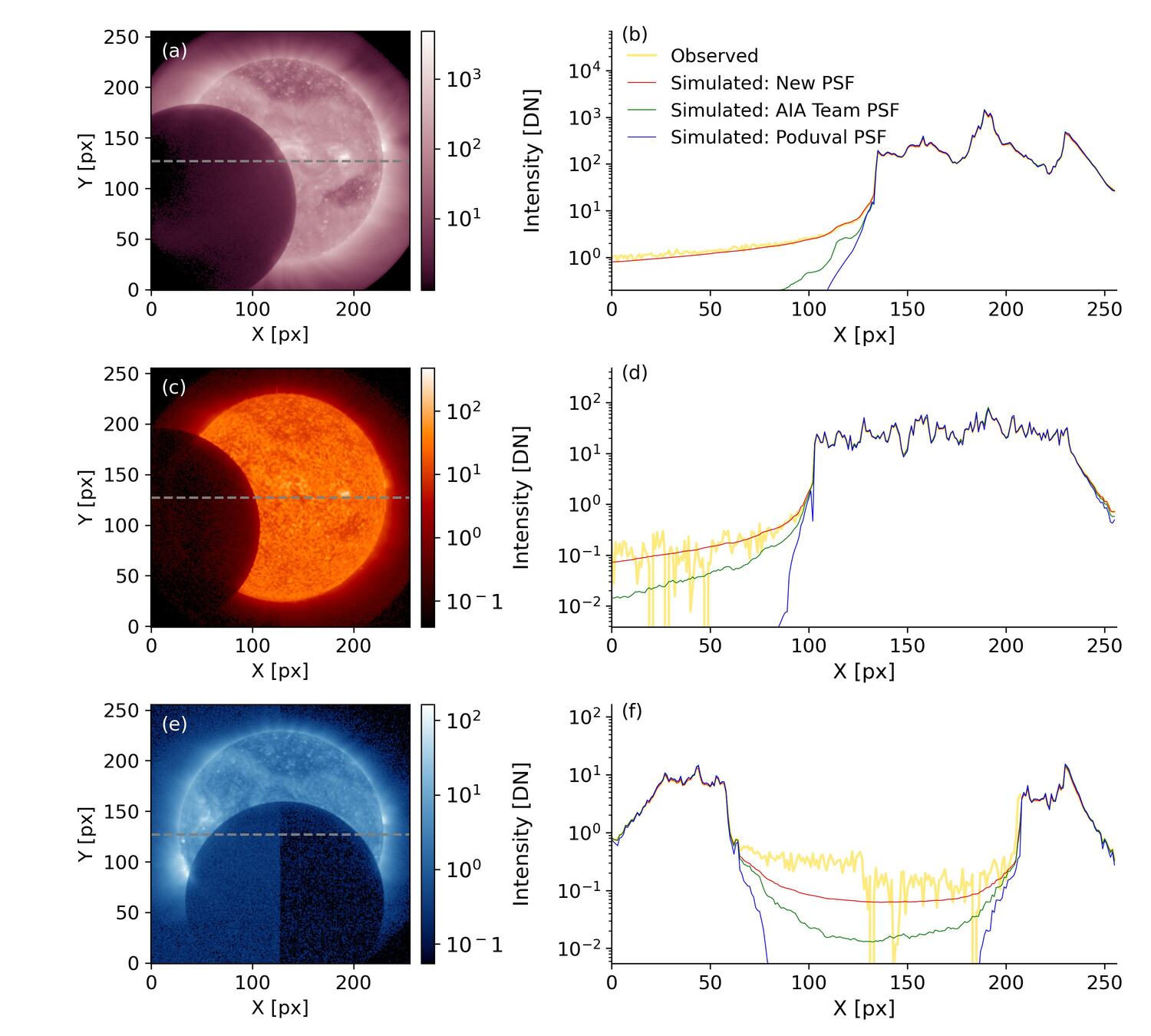}
    \caption{Continuation of Figure~\ref{fig:eval_eclipse2} for the AIA~$211$,~$304$, and~\SI{335}{\angstrom} channels.}
    \label{fig:eval_eclipse2}
\end{figure}

We evaluate the quality of our PSFs on the 2011~September~8 flare and on the 2016~October~30 lunar eclipse by comparing the image reconstructions with the ones from the AIA Team and those from \cite{poduval2013}.

In Figures~\ref{fig:eval_flares1} and~\ref{fig:eval_flares2}, we show the image reconstruction results for the 2011~September~8 flare. An accurate reconstruction would be indicated by a complete disappearance of the diffraction pattern. With our new PSFs, the residual artifacts at the location of the diffraction pattern are less intense than with the AIA Team or Poduval PSFs. This results because we have individually fitted for the mesh parameters for the horizontal and vertical mesh directions. However, the remaining residual artifacts suggest a slight residual misalignment between our derived PSF diffraction pattern and the observed flare diffraction pattern. The cause could either be uncertainties in the fitted PSFs; optical aberrations, i.e., the PSF is not constant over the AIA field of view; or that the PSFs vary slightly with time. Time variation might occur due to entrance mesh temperature variations and orbital centrifugal forces. However, as long as only the exact location of the diffraction peaks slightly changes, but not the amount of diffracted light, this residual misalignment has little effect for AIA image reconstructions. We note also that the correct reconstruction of flare intensities only depends on the PSF core and is, as such, independent of the reconstruction of the associated diffraction pattern  \citep{hofmeister2023}. 

Next, we evaluate the amount of medium- to long-distance scattered light in lunar-occulted regions. Often, this evaluation is done by deconvolving partially occulted images with the PSF and assessing how much of the light scattered has been removed from the occulted region. This approach, however, does not provide an accurate evaluation, since an incorrectly overestimated PSF would apparently remove all of the scattered light. Instead, we use the fitted PSF to derive the expected amount of light that is scattered into the occulted region, and compare these expected intensities with the observed intensities. To derive the expected amount of light scattered into the occulted region, we first reconstruct the illuminated portion of the image by deconvolving the image with the desired PSF. Then, we set the intensities in the occulted region to zero; this gives us the best approximation of the true image. Afterwards, we convolve this approximation of the true image again with the desired PSF to simulate the instrumental scattering. Finally, we compare the simulated intensities from the instrumental scattering in the occulted region with that observed.  We note that the average reconstructed true image on large scales is quite robust to errors in the PSF; that is, the quality of the reconstructed true image has only a minor effect on the amount of simulated scattered light. Any larger differences between the simulated and observed intensities in the occulted region can be attributed to errors in the scattering function, i.e., errors in the fitted PSF weights.

The results are shown in Figures~\ref{fig:eval_eclipse1} and~\ref{fig:eval_eclipse2}. Each row shows the 2016~October~30 lunar eclipse in one AIA channel, where we rebinned the images to a resolution of $256 \times 256$~pixels to reduce photon noise. Using our PSF, the simulated intensities in the occulted region closely follow the trend in the observed intensities, with typical deviations of \SI{0.3}{DNs}. The remaining difference between the simulated and observed intensities can be attributed to two effects: photon noise (visible in the AIA~$94$,~$304$, and~\SI{335}{\angstrom} panels) and calibration errors in the readout amplifiers between the four quadrants of the AIA CCD (visible as jumps in the intensities at the vertical center line in the~$94$,~$131$, and~\SI{335}{\angstrom} panels). We note that in the low-count \SI{335}{\angstrom} image, the calibration error relative to the mean intensity is large. This results in a small  mismatch between the observed and simulated intensities, the consequences of which will be further explored in the next section. Lastly, when using the PSFs of the AIA Team and of Poduval, the simulated intensities in the occulted region are too low by more than an order of magnitude, showing that they have significantly underestimated the long-distance scattered light. 

\subsection{Effects on AIA images and associated DEM analyses}

\begin{figure}
    \centering
    \includegraphics[width=1.\linewidth]{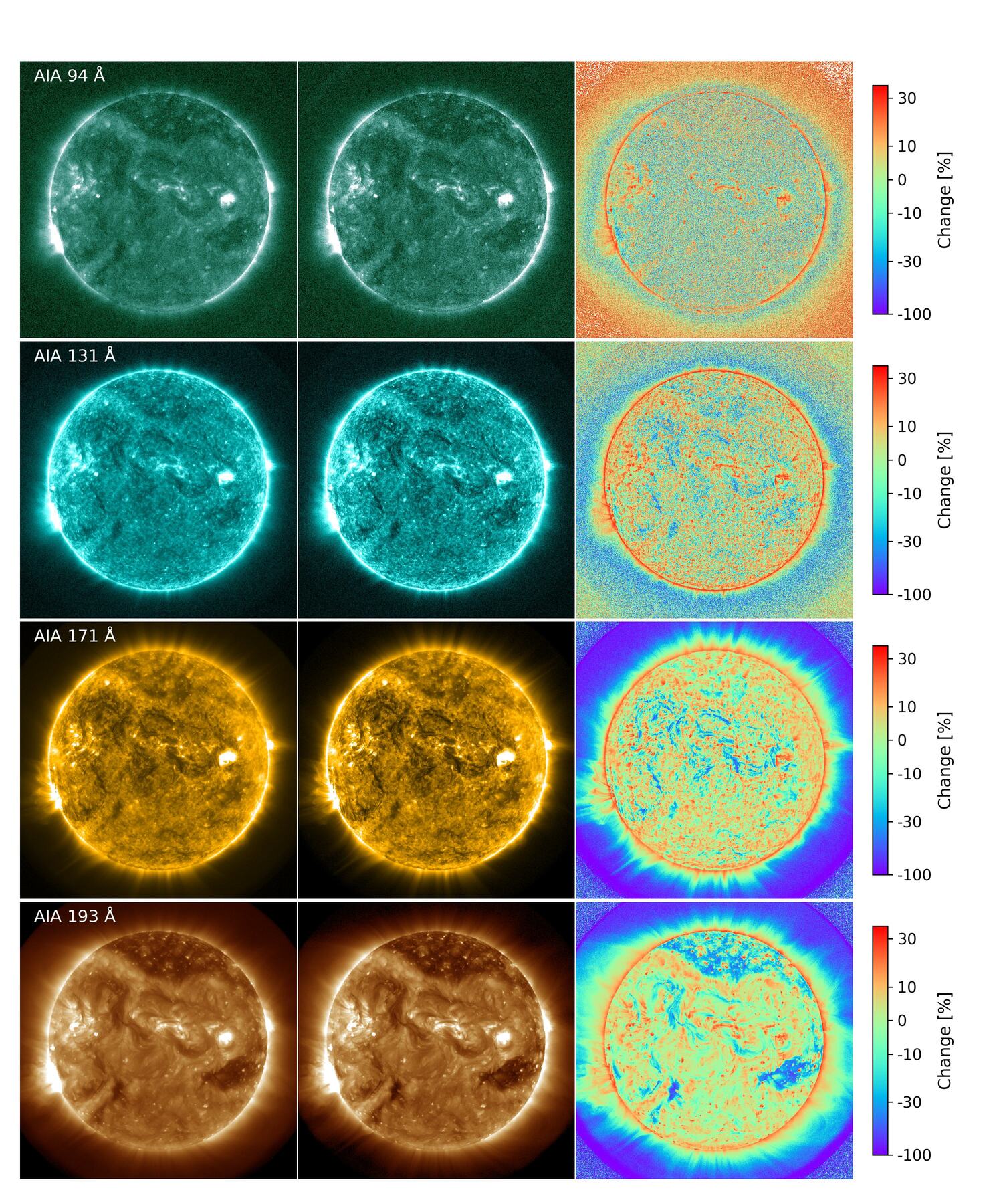}
    \caption{Effect of the PSF deconvolution on the AIA images of 2016~October~30. Left: original image. Center: Deconvolved image using the revised PSFs. Right: Percentage change from the original to the deconvolved image. From top to bottom: AIA~$94$,~$131$, $171$,~and~\SI{193}{\angstrom} images. }
    \label{fig:eval_imchange1}
\end{figure}
\begin{figure}
    \centering
    \includegraphics[width=1.\linewidth]{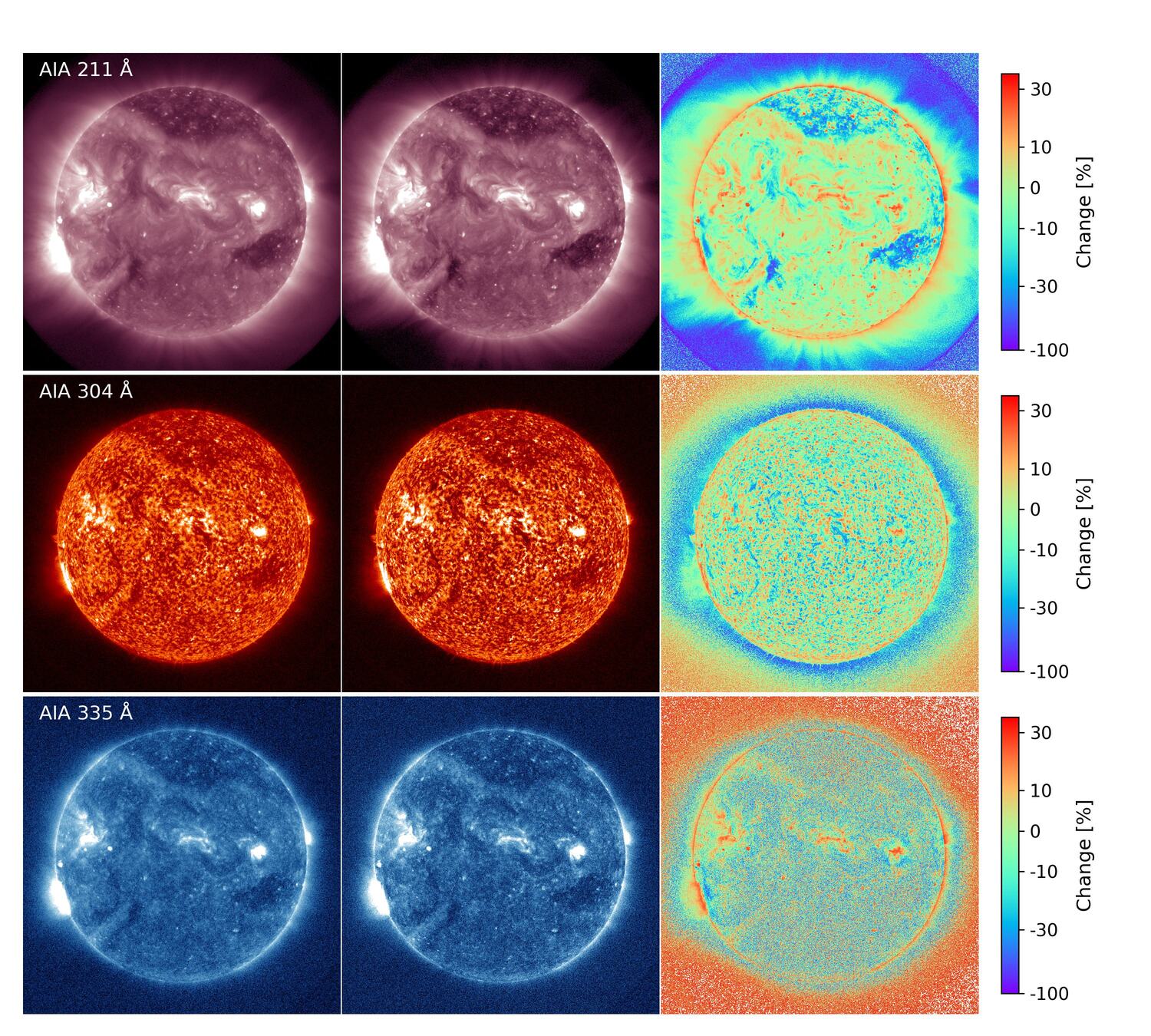}
    \caption{Continuation of Figure~\ref{fig:eval_imchange1} for the AIA~$211$,~$304$, and \SI{335}{\angstrom} channels.}
    \label{fig:eval_imchange2}
\end{figure}

\begin{figure}
    \centering
    \includegraphics[width=1.\linewidth]{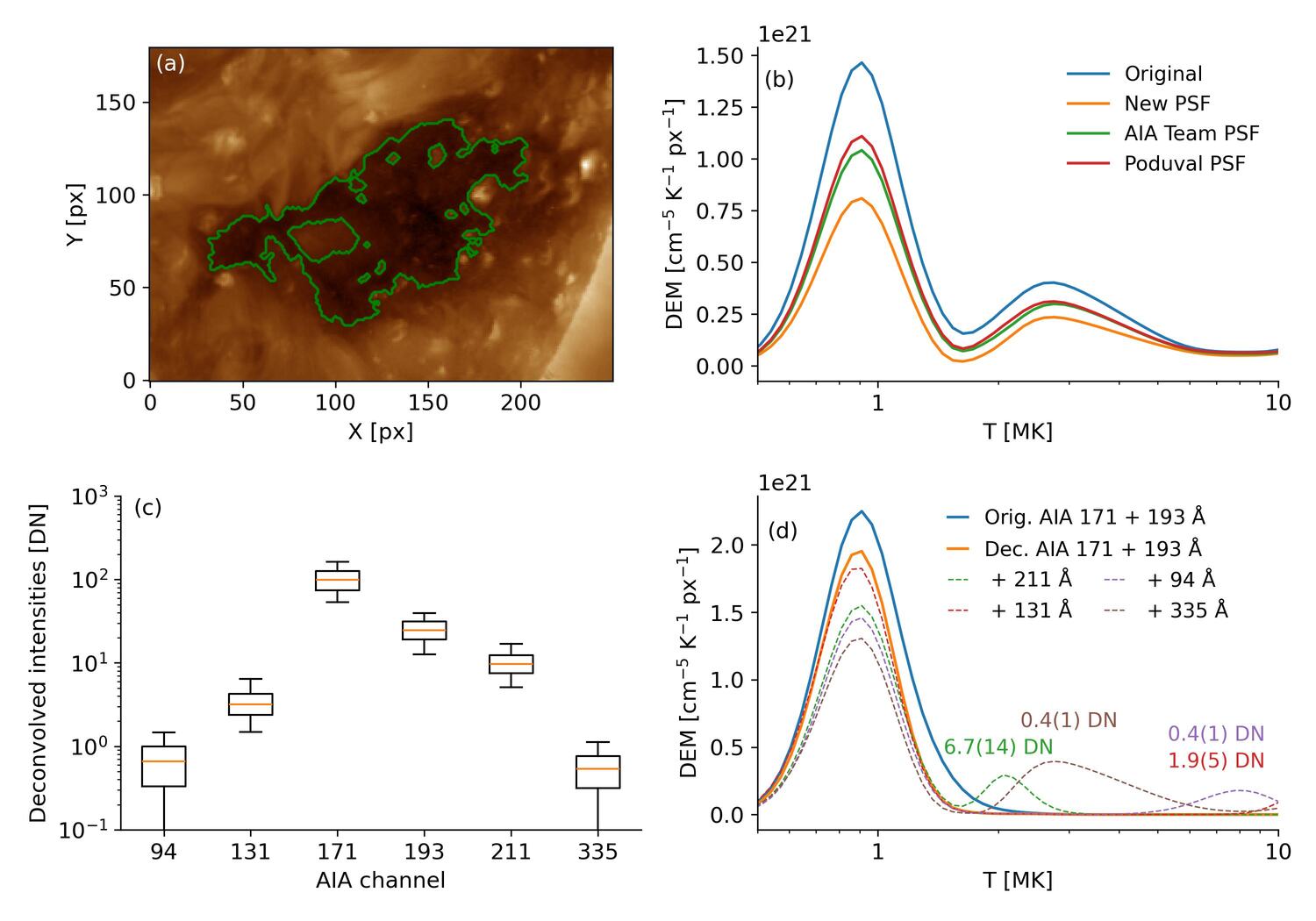}
    \caption{Effect of the PSF deconvolution on the DEM reconstruction of a coronal hole. (a) Image of the coronal hole taken on 2016~October~30. The green contour outlines the region considered. (b) DEM reconstruction using the AIA~$94$,~$131$, $171$,~$193$, $211$,~and~\SI{335}{\angstrom} channels as input. The blue line shows the DEM reconstruction from the original AIA images. The yellow, green, and red lines show the DEM reconstruction from the PSF-deconvolved images using our new, the AIA team, and the Poduval PSFs, respectively. (c) Observed counts of the coronal hole in each AIA channel.  For each channel, the orange line gives the median value, the box the $1\sigma$ range, and the extended error bars the $2\sigma$ range. (d) DEM reconstruction using only selected AIA channels. The blue line shows the reconstruction using the original AIA~$171$ and~\SI{193}{\angstrom} channels. The orange line uses the deconvolved AIA~$171$ and~\SI{193}{\angstrom} channels with our new PSF as input. The green, red, purple, and brown dashed lines, respectively, individually add the deconvolved~$211$, $131$,~$94$, and~\SI{335}{\angstrom} images to the deconvolved AIA~$171$ and~\SI{193}{\angstrom} DEM reconstruction. The numbers above the peaks give the corresponding AIA counts that cause these peaks. }
    \label{fig:dem_ch}
\end{figure}
\begin{figure}
    \centering
    \includegraphics[width=1.\linewidth]{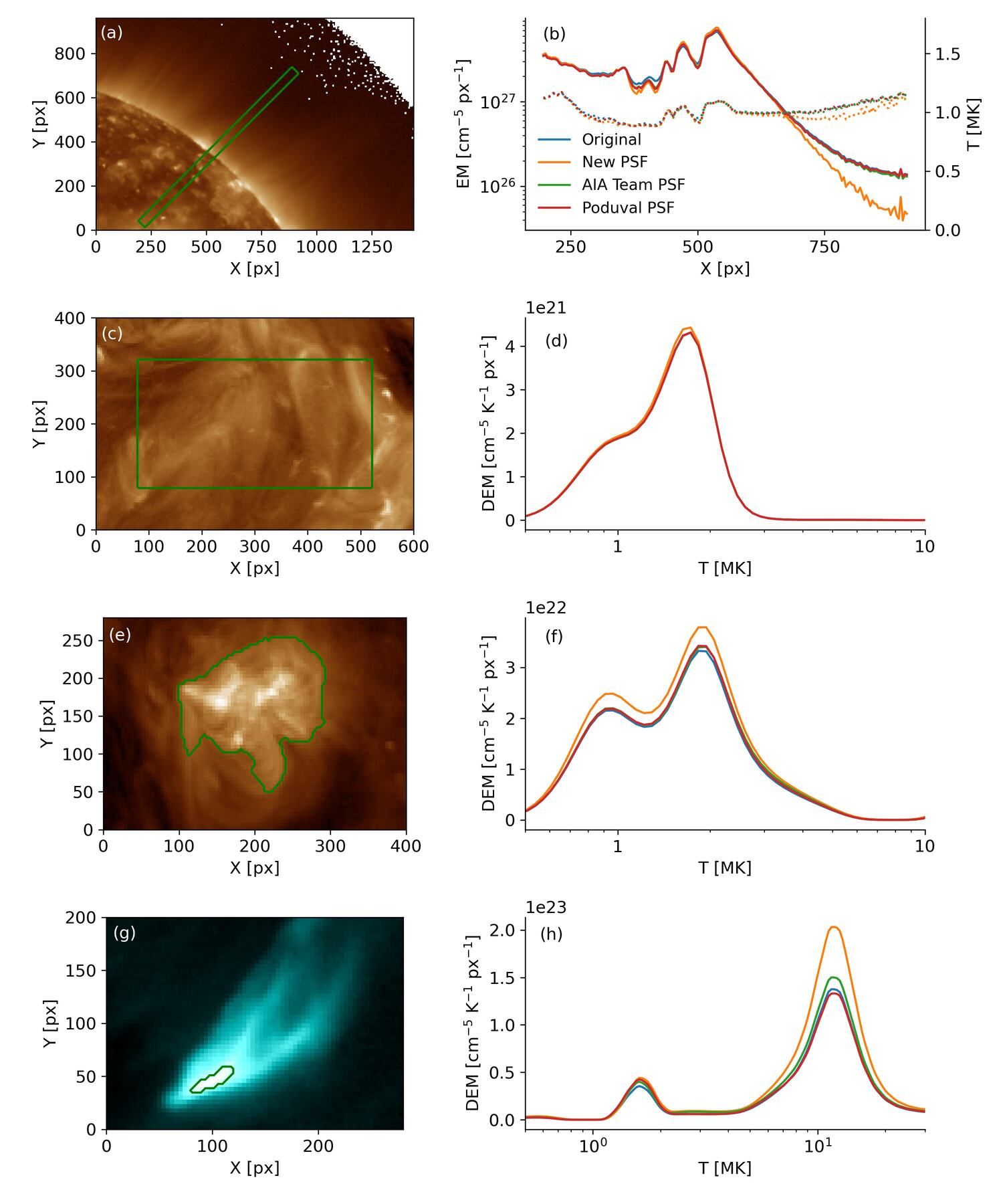}
    \caption{DEM reconstructions for various regions. First row: 2016~October~30 on-disk to off-limb region, reconstructed from the AIA~$171$ and \SI{193}{\angstrom} channels. The solid and dashed lines in the right panel give the EM and temperature evolution, respectively, within the green box in the left panel. Second row: 2016~October~30 quiet Sun region, reconstructed from the AIA~$171$,~$193$, and~\SI{211}{\angstrom} channels. The DEM is the average for the region outlined in green in the left panel. Third row:  2016~October~30 active region, reconstructed from all 6~coronal AIA channels. The DEM is the average for the region outlined in green in the left panel. Fourth row: 2022~April~20 flare, reconstructed from all 6~coronal AIA channels.  The DEM is the average for the region outlined in green in the left panel. The blue lines give the DEM reconstructions from the original images. The orange, green, and red lines reconstruct the DEM from the deconvolved images using our new, the AIA team, and the Poduval PSFs, respectively.}
    \label{fig:dem_others}
\end{figure}

Scattered light reduces the intensity in bright features and increases the intensity in dark regions of images. This results in a smaller dynamic contrast and might also affect the determination of physical properties, such as the amount and temperature of solar plasma as derived from a differential emission measure (DEM) analysis. In the following, we analyze the effects of the PSF deconvolution with our revised PSFs on solar images and DEM analyses. 

\paragraph{Solar images} In Figure~\ref{fig:eval_imchange1} and~\ref{fig:eval_imchange2}, we show the Sun as observed by AIA 10 minutes before the start of the 2016~October~30 lunar eclipse. A visual comparison of the original and deconvolved images reveals a slightly enhanced contrast, which is most visible in the AIA~\SI{171}{\angstrom} channel. When analyzing the percentage intensity change, shown in the right columns of the figure, it becomes apparent that bright features (such as active regions, coronal bright points, and bright loops) become brighter by up to \SI{30}{\percent}, while dark features (such as coronal holes in the AIA~$193$ and~\SI{211}{\angstrom} channels) become darker by up to \SI{90}{\percent}. 

The off-limb intensity also changes. In the~$171$,~$193$, and~\SI{211}{\angstrom} channels, the off-limb intensity in the PSF-deconvolved images decreases. This shows that a large portion of the off-limb intensity in the original images consisted of instrumental scattered light. In the~$94$ and~\SI{335}{\angstrom} channels, in contrast, the off-limb intensity decreases only close to the solar disk but increases everywhere else. This increase is a residual effect of the instrumental calibration combined with PSF deconvolutions for low-count channels. The measured off-limb intensity consists of real emission, instrumental scattered light from the solar disk, and an instrumental offset from the CCD amplifier calibration. In these low-count channels, the measured off-limb plasma emission is close to zero and the absolute amount of scattered photons from the solar disk to the off-limb regions is also close to zero, so that the instrumental offset becomes comparable to the measured off-limb counts. If the instrumental offset were real plasma emission, AIA would scatter part of its intensity outside the field of view. Deconvolving such an image with the PSF retrieves these wrongly assumed scattered photons back into the image, causing the artificial increase of off-limb intensities in the low-count AIA channels. 

\paragraph{DEM analyses} The solar corona consists of a highly ionized plasma, where the ionization state depends on the local coronal temperature.  By observing the emission of specific ions, AIA images the coronal plasma distribution at various temperatures. A DEM analysis aims to derive the original plasma temperature distribution from these images. This inversion problem is in general ill-defined but can be approximated by assuming additional constraints. We use the inversion method of \cite{hannah2012}, which solves this problem under the constraint of minimizing the amount of plasma that is needed to explain the observations.

In Figure~\ref{fig:dem_ch}, we show the DEM of a coronal hole observed on 2016~October~30. The DEM shows two peaks, one at about \SI{0.8}{MK} and a second one at about \SI{3}{MK}. As coronal holes are comparably cool features, this second peak is assumed to be an artifact of instrumental diffracted and scattered light from hotter coronal regions, such as active regions. Therefore, we assumed that deconvolving the image with the revised PSF should diminish the second peak. Although it is reduced, the deconvolution with the revised PSFs did not remove this second peak. This implies that there is another cause for the artifact. The second peak might be caused by instrumental offsets and measurement uncertainties in the low-count channels of AIA. Figure~\ref{fig:dem_ch} (c) shows that the average counts in the low-count channels are only~\SI{0.7}{DN} for AIA~\SI{94}{\angstrom}, \SI{3.5}{DN}~for~AIA~\SI{131}{\angstrom}, \SI{10}{DN}~for~AIA~\SI{211}{\angstrom}, and \SI{0.6}{DN}~for~AIA~\SI{335}{\angstrom}. To test this hypothesis, we first derive the DEM only using the comparably high-count AIA~$171$ and~\SI{193}{\angstrom} channels with average counts in the coronal hole of~\SI{102}{DN} and~\SI{25}{DN}, respectively. Then, we add each low-count channel individually to the AIA~$171$+$193$~\si{\angstrom} DEM and observe how the DEM changes. The result, presented in Figure~\ref{fig:dem_ch}~(d), shows that each low-count channel generates a DEM peak of its own.

Next, we estimate how many AIA counts in the low-intensity channels are required to generate these additional peaks. We begin with the AIA~$171$+$193$~\si{\angstrom} DEM, which we assume to be robust. Then, we calculate the number of counts in the low-intensity channels that are already accounted for by this DEM. These are given by the AIA~$171$+\SI{193}{\angstrom} DEM multiplied by the AIA contribution functions for the low-intensity channel being considered. The difference between the observed counts in the low-intensity channel and the counts in that channel due to the AIA~$171$+\SI{193}{\angstrom} DEM produces  the additional peaks in the DEMs that include a low-intensity channel, as shown in Figure~\ref{fig:dem_ch}~(d). We also note that the peak at about \SI{3}{MK} in our DEM is mainly an artifact created from only \SI{0.4}{DNs} in the \SI{335}{\angstrom} channel. These \SI{0.4}{DNs} are smaller than the uncertainty in the measurements,  which is about \SI{0.2}{DNs} for the read-out amplifier calibration plus \SI{0.6}{DNs} from the photon noise. This shows that instrumental and measurement effects in low-count channels can create significant artifacts in DEMs. Therefore, one should take great care when including channels with low-counts in DEM reconstructions. 

In Figure~\ref{fig:dem_others}, we present the effect of the PSF deconvolutions on an on-disk to off-limb DEM, a quiet Sun DEM, and an active region DEM derived from the AIA images taken on 2016~October~30; and a DEM of the 2022~April~20 flare.  Panels~(a) and~(b) present the emission measure (EM) and temperature evolution along an on-disk to off-limb line close to the northern polar coronal hole. For each image location, the EM is defined as the integral of the DEM and gives a measure on the total amount of plasma along the line of sight. The temperature here is the DEM-weighted average temperature. The DEM was derived from only the AIA~$171$ and~\SI{193}{\angstrom} channels to reduce artifacts from the low-count channels far off-limb. We find that the EM-curves derived from the AIA team PSF-deconvolved images and the Poduval PSF-deconvolved images follow the same off-limb trend as the EM-curve derived from the original images. In contrast, since our PSF describes better the long-distance scattered light, the EM-curve derived from our new PSF decreases much faster. Furthermore, when using our revised PSF, we obtain slightly lower off-limb temperatures. 

Figures~\ref{fig:dem_others}(c) and~(d) show a quiet Sun DEM averaged over the field of view outlined in green. This DEM was derived from the AIA~$171$,~$193$, and~\SI{211}{\angstrom} channels. The remaining AIA channels were in the noise regime. In this DEM,  instrumental diffracted and scattered light has no significant effect on the DEM reconstruction. The reason is that the intensity gradient over the image region is small, so that the number of photons that are lost at a location due to diffraction and scattering and the number of photons correspondingly gained from the surrounding region are on the average equal \citep{hofmeister2023}. Thus, when one averages over a sufficiently large area of similar intensities, such as the quiet Sun region shown here, instrumental diffracted and scattered light does not have a significant effect on DEM results.

Figures~\ref{fig:dem_others}(e) and~(f) show the DEM results for an active region. Here, we used all optically thin AIA channels, i.e., AIA~$94$,~$131$, $171$,~$193$, $211$,~and AIA~\SI{335}{\angstrom}, to derive the DEM. The DEM curves derived from the original images and those derived from the AIA team and Poduval PSF-deconvolved images are all close to each other. However, the DEM curve using our new PSFs increases the DEM peaks by up to \SI{20}{\percent}. This increase is a result of the retrieval of long-distance scattered photons. Similar increases can also be expected for smaller bright regions, such as coronal bright points or bright coronal loops. 

Figures~\ref{fig:dem_others}(g) and~(h) show the DEM results of the 2022~April~20 X~2.25~class flare. This DEM was derived from unsaturated short-exposure images at 3:52~UT, which was $5$~minutes before the  maximum intensity of the flare. The image deconvolution with our new PSFs increases the height of the DEM flare peak by about \SI{50}{\percent}, and slightly increases the peak temperature of the flare from \SI{11.5}{MK} to \SI{11.8}{MK}. The reason for the large increase in the DEM value is twofold: (1) the retrieval of medium- to long-distance diffracted and scattered photons and (2) the Hannah-Kontar DEM inversion tries to minimize the amount of plasma to explain the observations and favors here a solution with more plasma. Similar increases can also  be expected for the energy estimation for microflares and picoflares. 

These results show that an accurate consideration of medium- to long-distance diffracted and scattered photons enhances the dynamic contrast in the images. Furthermore, it can affect the results of calculations based on these images, such as DEM analyses. The strength of the effect depends on the brightness and size of the target of interest in relation to the average brightness of the rest of the image.

\section{Summary and Discussion}

We have revised the PSFs of AIA, where we focused on determining accurately the diffracted light and the diffuse scattered light. We found that:
\begin{itemize}
    \item The amount of light diffracted by the meshes is $27$~to~\SI{34}{\percent} and the amount of diffuse scattered light is $10$~to~\SI{35}{\percent}.    
    \item The AIA meshes, which generate the diffraction pattern, cannot be described by a single set of mesh parameters and have to be treated individually.
    \item Diffuse scattering has to be considered over the full size of the detector and can be described for AIA by the sum of two power laws. 
    \item We attempted to fit the PSF core but found that a precise determination of the core size is not possible with the current available data and methods. We were able to determine an upper limit on the Gaussian width of the PSF core of \SI{.6}{px}.
    \item Deconvolving AIA images with the revised PSFs increases the intensity of bright image features by \SI{\approx 30}{\percent} and reduces the intensity of dark image features by up to \SI{90}{\percent}.
    \item Using the revised PSFs constrains the DEM analysis of coronal holes to lower temperatures and increases the EM of flares by \SI{\approx 50}{\percent}, as derived by the Hannah-Kontar method.
\end{itemize}

We believe that the parameters of the AIA meshes, which generate the diffraction pattern, slightly change over time, probably due to mechanical and temperature deformations. Variations in the mesh parameters  have a negligible effect on the amount of diffracted light  and a small effect on the expected location of the diffraction peaks. Since the variations are small, this uncertainty will only have a very minor effect on the quality of image reconstructions by PSF deconvolutions. However, the uncertainty in the location of the diffraction peaks hinders attempts that try to estimate the intensity of saturated flare pixels from the observed diffraction pattern, such that of \citet{schwartz2015}. 

The main issues we encountered for the determination of the diffuse scattered light, which is scattered over the entire detector, were calibration artifacts that become apparent in the lunar-occulted regions. Although small in absolute counts, on the order of \SI{0.2}{DNs}, they  increase the uncertainty in the PSF determination, particularly for the low-count AIA~$94$, $131$, and~\SI{335}{\angstrom} channels. 

Determination of the PSF core proved to be the greatest challenge. We were able to determine an upper limit on the width of the PSF core by analyzing at which core width image artifacts start to appear. We believe that a precise determination of the PSF core width is not possible with the current data and methods at hand. Methods that attempt to estimate the width of the PSF core from the diffraction pattern in flare images suffer from the uncertainty in the location of the diffraction peaks. Methods that are based on the analysis of the intensities in the occulted regions suffer from calibration artifacts and the uncertainty in the amount of medium- to long-distance diffuse scattered light. Lastly, for blind deconvolution methods, we lack priors that are appropriate for observations of the optically thin solar corona. 
    
The accuracy of our PSFs depends on the condition of the AIA instrument and its calibration. Our analysis specifically assumes that each PSF remains shift invariant across the entire image, that the calibration of the AIA readout amplifiers is statistically accurate across all images utilized, and that no significant light leaks are present. A theoretical PSF accounting for diffuse scattered light, to which we could compare, is not currently available for AIA. However, \citet{martinez2010} provided a theoretical PSF for the Solar Ultraviolet Imager (SUVI), an EUV imager observing the Sun at similar wavelengths to AIA. Their calculations suggest that approximately 48--55\% of the light is diffracted and diffusely scattered, with 28--35\% extending beyond \SI{10}{\arcsec} and 3--9\% beyond \SI{100}{\arcsec}. Our results for AIA indicate a total diffraction and diffuse scattering of 37--55\%, with 20--25\% extending beyond \SI{10}{\arcsec} and 11--15\% beyond \SI{100}{\arcsec}. Thus, the PSF of AIA, as determined in our study, decreases at large scattering distances more gradually with distance from the PSF center than the theoretical PSF for SUVI. Nevertheless, considering the design differences between the instruments, the results are reasonably consistent.

For future instruments, we recommend measuring the PSF core width while the instrument is still on ground. Furthermore, we recommend taking extremely long exposure images of lunar occultations once the instrument is in space. We expect that the uncertainty in the amount of medium- to long-distance diffuse scattered light diminishes once the recorded intensities in the lunar-occulted region are far greater than the instrumental calibration uncertainties. Also, if it is possible to make the uncertainty of the diffuse scattered light sufficiently small, it might become possible to perform an in-flight characterization of the PSF core.

\ \\[.3cm]
\textit{\large Acknowledgement:} We thank the AIA team for the discussions during the course of this project. The AIA data is available by courtesy of NASA/SDO and the AIA, EVE, and HMI science teams.


\facility{SDO(AIA)}
\software{Astropy \citep{astropy1,astropy2,astropy3},  
    Matplotlib \citep{matplotlib},
    Numba \citep{numba},  
    Numpy \citep{numpy}, 
    Scikit-image\citep{scikit},
    Scipy \citep{scipy}, 
    Sunpy \citep{sunpy}}

\bibliographystyle{aa} 
\bibliography{bibliography}

\end{document}